\documentclass{article}

\usepackage{listings}
\usepackage{sectsty}
\usepackage{graphicx} 
\usepackage{float}
\usepackage{array}
\usepackage{listings}
\usepackage{minted}
\usepackage[table]{xcolor}

\subsubsectionfont{\normalfont}

\usepackage{microtype}
\usepackage{graphicx}
\usepackage{subcaption}
\usepackage{booktabs} % for professional tables
\usepackage{natbib}
\usepackage{hyperref}

\usepackage[preprint]{icml2026}

\usepackage{amsmath}
\usepackage{amssymb}
\usepackage{mathtools}
\usepackage{amsthm}

\usepackage[capitalize,noabbrev]{cleveref}

\theoremstyle{plain}

\theoremstyle{definition}

\theoremstyle{remark}

\usepackage[textsize=tiny]{todonotes}

\usepackage{xspace}
\usepackage{xcolor}
\usepackage{color}
\usepackage{soul}
\usepackage{minted}
\usepackage{tcolorbox}
\usepackage{inconsolata}
\usepackage{enumitem}
\tcbuselibrary{breakable}

\definecolor{stef-color}{rgb}{1, 0.898, 0.706}

\newcommand{\scirpy}{SCIRPy\xspace}
\newcommand{\python}{\textsc{Python}\xspace}
\newcommand{\dias}{\textsc{Dias}\xspace}
\newcommand{\modin}{\textsc{Modin}\xspace}
\newcommand{\dask}{\textsc{Dask}\xspace}
\newcommand{\koalas}{\textsc{Koalas}\xspace}
\newcommand{\pandas}{\textsc{Pandas}\xspace}
\newcommand{\pandasbench}{\textsc{PandasBench}\xspace}
\newcommand{\framework}{\textsc{RuleFlow}\xspace}

\newcommand{\snippetgen}{\textsc{SnippetGen}\xspace}
\newcommand{\rulegen}{\textsc{RuleGen}\xspace}
\newcommand{\codegen}{\textsc{CodeGen}\xspace}

\newcommand{\snippetgenfunc}{\ensuremath{\mathcal{S}}\xspace}
\newcommand{\rewritecomp}{\ensuremath{\mathcal{R}}\xspace}
\newcommand{\match}{\ensuremath{\mathcal{M}}\xspace}
\newcommand{\cell}{\ensuremath{\psi}}
\newcommand{\cells}{\ensuremath{\Psi}\xspace}

\newcommand{\pairs}{\ensuremath{P}\xspace}
\newcommand{\rrule}{\ensuremath{r}\xspace}
\newcommand{\rrules}{\ensuremath{R}\xspace}
\newcommand{\snippetprompt}{\ensuremath{\pi_\snippetgenfunc}}

\newcommand{\runtime}{\text{runtime}}
\newcommand{\threshold}{\tau_r}
\newcommand{\Threshold}{\tau_a}

\newcommand{\vars}{V}

\newcommand{\constants}{C}

\newcommand{\identifiers}{I}

\definecolor{iterblue}{HTML}{3399FF}

\definecolor{itergray}{HTML}{8B6B9A}

\definecolor{iterpurple}{HTML}{990099}

\definecolor{backcolour}{RGB}{250, 250, 250}

\definecolor{keywordcolor}{rgb}{0.05,0.5,0.9}
\definecolor{asttype}{rgb}{0.05,0.05,0.9}   

\newcommand{\highlightvar}[1]{\bfseries\textcolor{keywordcolor}{#1}}
\newcommand{\highlightast}[1]{{\textcolor{asttype}{#1}}}

\usepackage{graphicx}
\usepackage{booktabs}
\usepackage{caption}
\usepackage{xcolor}

\icmltitlerunning{\framework: Generating Reusable Program Optimizations with LLMs}

\begin{document}

\twocolumn[
  \icmltitle{\framework: Generating Reusable Program Optimizations with LLMs}

  \icmlsetsymbol{equal}{*}

  \begin{icmlauthorlist}
    \icmlauthor{Avaljot Singh}{equal,yyy}
    \icmlauthor{Dushyant Bharadwaj}{equal,yyy}
    \icmlauthor{Stefanos Baziotis}{yyy}
    \icmlauthor{Kaushik Varadharajan}{yyy}
    \icmlauthor{Charith Mendis}{yyy}
  \end{icmlauthorlist}

  \icmlaffiliation{yyy}{University of Illinois Urbana-Champaign, USA}

  \icmlcorrespondingauthor{Avaljot Singh}{avaljot2@illinois.edu}
  \icmlcorrespondingauthor{Dushyant Bharadwaj}{db50@illinois.edu}
  \icmlkeywords{Machine Learning, ICML}

  \vskip 0.3in
]

\printAffiliationsAndNotice{\icmlEqualContribution}

\begin{abstract}
    Optimizing \pandas programs is a challenging problem. 
    Existing systems and compiler-based approaches offer reliability but are either heavyweight or support only a limited set of optimizations. Conversely, using LLMs in a per-program optimization methodology can synthesize nontrivial optimizations, but is unreliable, expensive, and offers a low yield. In this work, we introduce a hybrid approach that works in a 3-stage manner that decouples dicovery from deployment and connects them via a novel bridge. First, it discovers per-program optimizations (\textit{discovery}). Second, they are converted into generalised rewrite rules (\textit{bridge}). Finally, these rules are incorporated into a compiler that can automatically apply them wherever applicable, eliminating repeated reliance on LLMs (\textit{deployment}).  
    We demonstrate that \framework is the new state-of-the-art (SOTA) \pandas optimization framework on \pandasbench, a challenging \pandas benchmark consisting of \python notebooks. Across these notebooks, we achieve a speedup of up to $4.3\times$ over \dias, the previous compiler-based SOTA, and $1914.9\times$ over \modin, the previous systems-based SOTA. 
    % We also show that LLM-generated rewrite rules achieve a speedup of up to $205\times$ on individual cells compared to \dias and up to $1460\times$ compared to \modin.

    Our code is available at \url{https://github.com/ADAPT-uiuc/RuleFlow}.
\end{abstract}

\section{Introduction}
\label{sec:intro}
\pandas is a widely-used library for exploratory data analysis (EDA) workloads in both industry and research~\cite{modin,pandasFoundation,pandasbench}. Its intuitive, dataframe-centric API makes it popular in data science, finance, and scientific computing, spanning domains from exploratory analysis to feature engineering~\cite{whyPandas,pandasFinanceMedium,pandasForScientificComputing,pandasEmperical}. 
% Due to extensive \pandas usage, 
Several existing works have explored improving the performance or automation of \pandas programs through a variety of techniques that reduce the computational cost and the runtime of Pandas using code. However, most of these techniques underperformed when tested on a representative set of real-world benchmarks~\cite{pandasbench}.

\paragraph{Existing Techniques and Their Limitations.}
Prior work on optimizing \pandas can be broadly categorized into: (i) systems-based and (ii) compiler-based solutions. 
% Both face fundamental limitations.
% 
Systems-based solutions aim to provide broad API coverage~\cite{modin,dask,koalas,fireducks,polars}. However, they are often heavyweight, making them ill-suited for EDA workloads and frequently leading to slowdowns in practice~\cite{pandasbench,dias}. In contrast, compiler-based solutions~\cite{dias,scirpy} are lightweight, but they rely on a limited set of manually engineered patterns or rewrite rules. This reliance restricts their optimization scope and prevents them from delivering consistent speedups across representative EDA workloads~\cite{pandasbench}. Further, extending them to cover new patterns typically requires significant manual effort, limiting their evolvability and practical coverage.

\begin{figure}[]
    \centering
    \begin{tcolorbox}[colback=backcolour, colframe=white, boxrule=0.1pt, arc=0.4mm, left=0.1mm, right=0.5mm, top=0.1mm, bottom=0.1mm]
        \textbf{Original Code:} 
        \begin{minted}[fontsize=\small,breaklines,breakanywhere, escapeinside=||]{python}
df = df.drop(['name'], axis=1)
        \end{minted}
        \textbf{LLM-proposed Optimized Code:} 
        \begin{minted}[fontsize=\small,breaklines,breakanywhere, escapeinside=||]{python}
df.pop('name')
        \end{minted}
    \end{tcolorbox}

    \caption{Example optimization proposed by LLM. The optimized code is $1770\times$ faster on a representative dataframe size. 
    % The original code snippet and the dataframe are taken from KGTorrent ~\cite{kgtorrent}.
    }
    \label{fig:examplePair}
\end{figure}
\paragraph{Using LLMs.}
With the advent of large language models (LLMs), a complementary direction for optimizing \pandas becomes possible. A natural approach is to apply LLMs for \textit{per-program optimization}, i.e., prompting an LLM to optimize each \python code snippet that uses the \pandas API. While this approach discovers non-trivial optimizations that were not covered by any of the above systems (Figure~\ref{fig:examplePair}), it is difficult to scale in practice. LLMs incur nontrivial latency and monetary cost, and their outputs are often incorrect or fail to improve performance. As illustrated in Figure~\ref{fig:sankeySnippetgen}, when LLMs are prompted to optimize original code snippets, only $64\%$ (2,639 out of 4,138) of the proposed optimizations are semantically correct, and only $14\%$ (235 out of 2,639) of these correct optimizations yield performance improvements. This results in an overall yield of $5.7\%$. Consequently, for realistic \pandas programs, the probability of producing a valid optimization is low, rendering the per-program optimization approach unreliable. 

% \charith{So, we asked the question can we combine the benefits of traditional approaches and LLM-based optimizations to mitigate the problems of each other. More specifically, can we build an evolvable optimization framework with minimal runtime overhead? To answer this question, we specifically targeted Pandas-compilation systems.}

% \paragraph{Beyond \pandas.}
% \charith{the paragraph title sounds weird, the content is good; symmetry is missing. People may ask why do we consider only compilers? What is the use in systems solutions? Look at what I wrote above.} 
Optimizing \pandas programs exemplifies a broader challenge in designing program optimizations for real-world systems, where the space of possible optimizations is combinatorially large. Purely systems/compilers-based approaches offer reliability, but are heavyweight or support only a limited set of optimizations. In contrast, LLMs synthesize nontrivial optimizations; however, they are computationally expensive and unreliable. 
The central question, then, is: 
\begin{quote}
    \textit{Can we combine the benefits of traditional approaches and LLM-based optimizations, while avoiding their respective limitations?}
\end{quote}
% More specifically, can we build an evolvable optimization framework with minimal computational overhead? 

\paragraph{Key Insight: Low Yield, High Reuse.}
While successful optimizations are infrequent overall ($5.7\%$ in \pandas), recurring patterns often appear among them. So, the key idea is that such optimizations, once discovered, can be reused across many programs. In Figures~\ref{fig:notebookHitRate}, ~\ref{fig:ruleHitRate}, we show that if we reapply the discovered optimizations to unseen programs at test time, they collectively apply to $87.13\%$ of the total test notebooks, and collectively the notebooks use $18.68\%$ of the proposed optimizations. 
% So, a single high-quality optimization can be applied wherever it matches, amortizing discovery costs across many programs. 
This shift, from per-program optimizations to amplifying rare but high-quality optimizations by reusing them, is central to our approach. It amortizes the discovery cost and thus fundamentally changes how LLM-generated optimizations should be used.

\paragraph{Hybrid Approach.}
We propose a hybrid approach that separates \textit{discovery} from \textit{deployment}. During discovery, LLMs are used offline on a corpus of programs to explore possible optimizations. Each candidate optimization is automatically tested for correctness and performance improvements. During deployment, a lightweight, deterministic optimizer reuses these optimizations and does not rely on LLMs.
However, there is a challenge---the optimizations produced by LLMs are specific to the original code from which they were generated. For example, the optimized code in Figure~\ref{fig:examplePair} only works with the original code shown there; even small syntactic changes to it requires regenerating the optimization. So, LLM outputs cannot be directly deployed as general-purpose optimizations, and thus, bridging the development and deployment is non-trivial. To address this, each validated optimization is lifted from a concrete code instance into a generalized \textit{rewrite rule}.
\begin{figure}[]
    \centering
    \begin{tcolorbox}[colback=backcolour, colframe=white, boxrule=0.1pt, arc=0.4mm, left=0.1mm, right=0.5mm, top=0.1mm, bottom=0.1mm]
        \textbf{LHS:} 
        \begin{minted}[fontsize=\small,breaklines,breakanywhere, escapeinside=||]{python}
@{|\highlightast{Name}|: |\highlightvar{v1}|} = @{|\highlightast{Name}|: |\highlightvar{v1}|}.drop([@{|\highlightast{Const(str)}|: |\highlightvar{c1}|}], axis=1)
        \end{minted}
        \textbf{RHS:} 
        \begin{minted}[fontsize=\small,breaklines,breakanywhere, escapeinside=||]{python}
@{|\highlightvar{v1}|}.pop(@{|\highlightvar{c1}|})
        \end{minted}

        \textbf{Runtime Preconditions:}
        \begin{minted}[fontsize=\small,breaklines,breakanywhere, escapeinside=||]{python}
[isinstance(@{|\highlightvar{v1}|}, pandas.DataFrame), 
@{|\highlightvar{c1}|} in @{|\highlightvar{v1}|}.columns]
        \end{minted}
    \end{tcolorbox}

    \caption{This rule optimizes single column deletion by replacing \texttt{drop} with \texttt{pop}. This rewrite achieves a mean speedup of $18.31\times$ with a maximum speedup of $130.22\times$ on \pandasbench.}
    \label{fig:rewrite_rule}
\end{figure}

For this, we develop an LLM-based agent that analyzes the original code and its optimized version to identify generalizable components. 
It also identifies any semantic constraints on the applicability of the optimization. They are then encoded as rewrite rules in a domain-specific language (DSL), borrowed from \dias~\cite{dias}. The LLM-generated rewrite rule corresponding to the optimization in Figure~\ref{fig:examplePair} is shown in Figure~\ref{fig:rewrite_rule}. 
This bridge allows the optimization to apply to a broader class of syntactically and semantically similar programs, instead of a single program.

% \paragraph{Novelty.}
By separating speculative, high-variance discovery from deterministic, lightweight deployment, our design converts a low-yield process into a reliable and scalable optimization pipeline. While optimization-discovery and rule generation remain challenging, the resulting rewrite rules accumulate over time and amortize the cost of LLM usage, enabling performance improvements without introducing LLM overhead at the deployment stage.
Although we demonstrate our approach on \pandas, we believe that separating discovery from deployment, along with a bridge component, could offer a promising framework for other optimization venues.

% \charith{remove; repetition. I would add a small commentary to hypothesize the generality of our approach to many compiler problems. In pharsing like, "We believe ....}

% \paragraph{Novelty: Bridging LLMs and Compiler Techniques.}
% The key novelty of our approach lies in converting LLM-discovered program optimizations into reusable compiler rewrite rules. While prior work typically applies LLM outputs directly to individual programs, we instead treat LLM-generated optimizations as candidates for general rules that can be formalized, validated, and reused. Each rule captures a high-quality optimization together with its structural and semantic applicability conditions, enabling it to be checked once and then safely applied across many programs.
% By elevating individual LLM successes into persistent rewrite rules, our approach bridges the exploratory power of LLMs with the safety and efficiency of compiler-style optimization. This conversion transforms sporadic, low-yield LLM outputs into scalable and repeatable performance improvements, a capability that has not been explored in prior \pandas optimization systems.

% \charith{Out of place. Repetition.}
% A key characteristic of the \pandas API is its richness: many data optimizations can be expressed in multiple, semantically equivalent ways using different combinations of dataframe operations~\cite{pandasFoundation, dias}. As a result, functionally identical \pandas programs may exhibit substantially different performance, depending on the particular API choices made by the user.

\begin{figure*}[t]
    \centering
    \includegraphics[width=\linewidth]{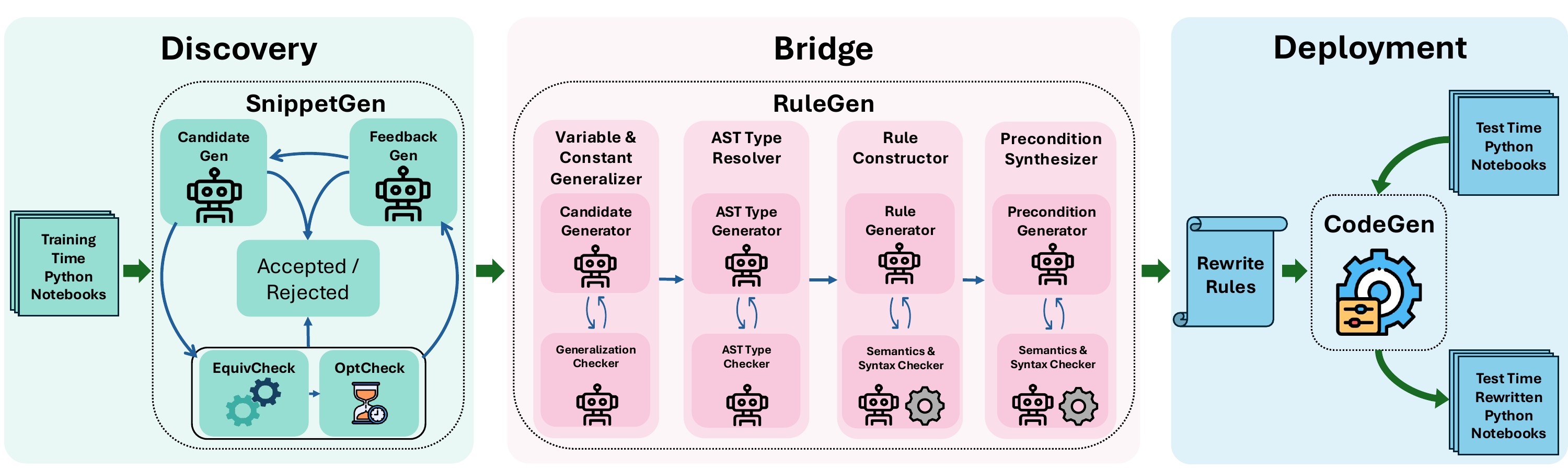}
    \caption{\framework is a \pandas optimization framework with three stages: (i) \snippetgen discovers candidate optimizations from code snippets, (ii) \rulegen converts these optimizations into rewrite rules, and (iii) \codegen applies the rules to unseen \pandas programs. The framework segregates LLM-based optimization discovery and compiler-based efficient rule deployment.}
    \label{fig:pipeline}
\end{figure*}

\paragraph{This Work.}
Building on our hybrid approach, we present \framework, a three-stage system that discovers optimizations, generates rewrite rules, and deploys them on real-world \pandas programs (Figure~\ref{fig:pipeline}). In \framework, LLMs identify optimizations and convert them into rewrite rules, which are then applied through a rewriter engine. 
Concretely, this paper makes the following contributions:
\begin{itemize}
    \item \textbf{Novel optimization strategy.} 
    We introduce a program optimization strategy that learns reusable rewrite rules, enabling amortized optimization across many programs rather than per-program optimizations.
    
    \item \textbf{End-to-end system.} 
    We implement \framework, a \pandas optimization framework that integrates LLM-driven discovery, automatic rule extraction, and deployment of rewrite rules to unseen notebooks. 

    \item \textbf{High-quality rewrite rules.} 
    \framework produces a set of high-quality \pandas rewrite rules that capture performance-critical optimizations.

    \item \textbf{Evaluation on real-world notebooks.} 
    We conduct extensive evaluation on \pandasbench~\cite{pandasbench}, a challenging benchmark comprising 102 real-world \python notebooks that use the \pandas API.
\end{itemize}

Overall, our evaluation demonstrates that \framework establishes a new state-of-the-art (SOTA) on \pandasbench. \framework outperforms the previous overall and compiler-based SOTA, \dias, up to $4.3 \times$, and exceeds the best systems-based approach, \modin, up to $1914.9\times$. At a finer granularity, our rewrite rules achieve speedups of up to $199 \times$ on individual cells compared to \dias and up to $1704 \times$ compared to \modin.
Further, our rules exhibit a high hit rate on \pandasbench. Individual rules apply to a maximum of $72$ notebooks, with individual notebooks matching upto $13$ distinct rules. 
These results confirm that converting LLM-discovered optimizations into reusable rewrite rules yields scalable, reliable performance gains that outperform existing \pandas optimization techniques.

\section{Related Work and Background}

We discuss the traditional \pandas optimization techniques (\S~\ref{sec:optbackground}), LLM-based program optimization techniques (\S~\ref{sec:kernelgen}), and the DSL used for rewrite rules (\S~\ref{sec:diasdsl}).

% In \S~\ref{sec:optbackground}, we describe \pandas optimization techniques. Next, in \S~\ref{sec:kernelgen}, we discuss related works that use LLMs for code optimization. Finally, in \S~\ref{sec:diasdsl}, we present the domain-specific language (DSL) borrowed from \cite{dias}, used to describe our rewrite rules.

\subsection{\pandas optimization techniques.}
\label{sec:optbackground}
% \paragraph{Rewriting} 
Rewriting techniques manipulate the user-written (\python) code, calling the \pandas API. By analyzing the code, they try to identify and improve sub-optimal code patterns. 
Two major works in this direction are \dias~\cite{dias} and \scirpy~\cite{scirpy}. \dias is a dynamic rewriter for Pandas code. It has a fixed set of 10 dynamic rewrite rules that it applies to the input \pandas code dynamically. 
\scirpy also uses rewriting, using static analysis to identify optimization opportunities.

% \paragraph{Pandas Replacements} 
Further, some approaches provide alternate implementations of the \pandas API,  improving some of the existing weaknesses, such as it is single-threaded, and it does not take advantage of the storage device~\cite{pandas_mem}. Several works address these issues, including \modin~\cite{modin}, \dask~\cite{dask}, \koalas~\cite{koalas}, FireDucks~\cite{fireducks}, cuDF~\cite{cudf}, PolaRS~\cite{polars}, Spark (with its DataFrame API~\cite{spark_sql}), and Vaex~\cite{vaex}. The problem with these is that they do not perform well on the \pandasbench, which focuses on EDA settings.

\subsection{LLM for Code Optimization}
\label{sec:kernelgen}

Recent work has explored the use of LLMs for code optimization. Most of these approaches use LLMs in a per-program setting, where the model directly optimizes a given program or kernel. Examples include AccelOpt~\cite{accelopt}, KernelBench~\cite{kernelbench}, AutoComp~\cite{autocomp}, ASPL~\cite{aspl}, Astra~\cite{astra}, ADRS~\cite{llmsql}, GenRewrite~\cite{genrewrite}, AutoTriton~\cite{autotriton}, OpenEvolve~\cite{openevolve},  AlphaEvolve~\cite{alphaevolve}, and R-Bot~\cite{sun2025rbotllmbasedqueryrewrite}, which apply LLMs to optimize a fixed set of programs or kernels. These methods can optimize individual programs, but they require repeated LLM calls and don’t generalize beyond the specific programs. Further, using LLMs directly at deployment (e.g., during compilation~\cite{llmcompiler}) produces incorrect outputs most of the time.

Also, to the best of our knowledge, there is no prior work that uses LLMs for \pandas optimization. However, some related works apply LLMs for \pandas program synthesis. For instance, AutoPandas~\cite{autopandas} and PandasAI~\cite{autopandas} generate \pandas programs using input-output examples or natural-language specifications. 
% LLMs have been used more broadly for code generation and rewriting tasks. Systems such as Clover~\cite{clover} and GenRewrite~\cite{genrewrite} also use LLMs to synthesize or transform code, demonstrating the expressive power of these models for program manipulation. However, these approaches similarly operate at the level of individual programs and rely on the LLM during each optimization or generation step.

In contrast, for the first time, we propose a novel approach to use LLMs for program optimization that focuses on discovering optimizations that generalize across programs. Rather than using LLMs directly at deployment time, we use them only in an offline discovery stage to generate candidate optimizations, which are then converted into reusable rewrite rules. These rules can be applied repeatedly by a compiler without further LLM interaction. This design avoids the cost and unreliability associated with per-program LLM optimization and mitigates the high error rates observed in prior attempts to use LLMs directly for compilation tasks.

\subsection{Domain-Specific Language (DSL) for Rewrite Rules}
\label{sec:diasdsl}

For rewrite rules, we use a subset of the DSL introduced in \dias~\cite{dias}. A rewrite rule consists of three parts: (i) LHS, which is the code pattern to match; (ii) RHS, which is the optimized code; and (iii) preconditions, the constraints under which the rule is applicable. An example rewrite rule is shown in Figure~\ref{fig:rewrite_rule}.
All components are valid \python codes, except for abstract variables. These variables consist of an AST type and the variable name. For example, \texttt{@\{\highlightast{Name}: \highlightvar{var}\}} matches an expression of type \texttt{\highlightast{Name}} (as defined in \python’s AST~\cite{ast}), assigning it the identifier \texttt{\highlightvar{var}}, which can then be referenced in the RHS or preconditions. In the rule in Figure~\ref{fig:rewrite_rule}, the expression \texttt{@\{\highlightast{Name}: \highlightvar{df}\}} is an abstract variable. The precondition requires \texttt{\highlightvar{df}} to be a dataframe. 
At runtime, any \python element of the \texttt{\highlightast{Name}} AST type will match \texttt{\highlightvar{df}}, and the rule will apply at runtime if the matched element is a dataframe.

\section{\framework}
\label{sec:technical}

To scale \pandas optimizations beyond what either systems/compiler-based methods or LLMs can achieve alone, we design a three-stage framework, \framework~(Figure~\ref{fig:pipeline}), which systematically converts LLM-discovered improvements into reusable, automatically applicable rewrite rules. The first stage, \snippetgen~(\S~\ref{sec:snippetgen}), takes real-world \pandas code, generates multiple candidate optimized variants, and tests them for correctness and performance improvements. The second stage, \rulegen~(\S~\ref{sec:rulegen}), generalizes each pair of original and optimized code into a parameterized rewrite rule that captures the underlying optimization pattern rather than the specific instance. Finally, the \codegen~stage (\S~\ref{sec:codegen}) is a lightweight compiler that applies the rewrite rules to unseen \pandas programs through static pattern matching, without further LLM calls. 
Together, these stages leverage LLMs for discovering transformations and compiler techniques for scalable rule-based deployment.

\subsection{Discovery Stage: \snippetgen}
\label{sec:snippetgen}
The goal of \snippetgen is to explore the space of \pandas optimizations to generate semantically equivalent but potentially faster rewrites of real-world code.

\paragraph{Problem Setup}
Let \cells denote the set of code cells extracted from a corpus of Jupyter notebooks. We model these cells as samples $\cell \sim \mathcal{D}_\cells$, drawn from an unknown distribution of real-world \pandas code snippets. 
% In our implementation, we collect a large corpus of Kaggle~\cite{} notebooks and treat each notebook cell independently. 
The aim is to propose candidate rewrites that are semantically equivalent but offer runtime improvements. We define a candidate generation function $\snippetgenfunc: \cells \rightarrow 2^\cells$, which maps an input snippet $c \in \cells$ to a set of candidate rewrites $\{\cell'_1, \cell'_2, \cdots\}$.

\paragraph{AI Agent for Candidate Generation.}
We implement the function $\snippetgenfunc$ using an LLM-assisted AI agent (Figure~\ref{fig:pipeline}). For each cell $\cell \in \cells$, the agent constructs a prompt $\snippetprompt(\cell)$ that instructs an LLM to propose optimized rewrites (Appendix~\ref{appendix:prompts}).  
The LLM, denoted $\mathcal{L}$, is treated as a stochastic function producing candidate rewrites 
\(\cell'_i \sim \mathcal{L}(\snippetprompt(\cell))\), with multiple candidates $(i = 1, \dots, k)$ sampled per cell. The resulting set of candidates is $\snippetgenfunc(\cell) = \{\cell'_1, \dots, \cell'_k\}.$
We refer to this step as \textbf{CandidateGen}. Each candidate $\cell'_i \in \snippetgenfunc(\cell)$ is then evaluated against two constraints:
\begin{enumerate}
    \item \textbf{EquivCheck Constraint ($\phi_e$).} The candidate must preserve the semantics of the original cell: 
    \(\phi_e(\cell'_i) \triangleq \cell'_i \equiv \cell\). Because \pandas and \python lack formal semantics, we approximate this check via testing: both programs are executed on random dataframes $D = \{d_1, \dots, d_m\}$, and equivalence is accepted if $\forall d_j, \cell'_i(d_j) = \cell(d_j)$.

    \item \textbf{OptCheck Constraint ($\phi_p$).} A candidate must yield a measurable runtime improvement. It is retained only if the average improvement exceeds a threshold. We define the absolute improvement as $\Delta_i = \runtime(\cell) - \runtime(\cell'_i) - \Threshold$ and the relative improvement as $\delta_i = \frac{\runtime(\cell) - \runtime(\cell'_i)}{\runtime(\cell)} - \threshold$. The candidate passes if \(\phi_p(\cell'_i) \triangleq \min(\Delta_i, \delta_i) \geq 0\). In our implementation, we set $\Threshold$ to $150$ ms and $\threshold$ to 2. 
\end{enumerate}

\paragraph{Feedback.}
Since equivalence is established via testing rather than formal verification, candidate rewrites may still fail on unseen inputs. To refine their quality, we introduce an adversarial validation step, \textbf{FeedbackGen}, which actively attempts to falsify candidate rewrites.  
If no counterexamples are found, the rewrite is accepted and passed to the next stage. Otherwise, the counterexamples are provided as structured feedback to the LLM-driven generator.
Upon receiving counterexamples, CandidateGen may respond in one of three ways:
(i) \emph{Refute the counterexamples} by generating suitable preconditions in the \rulegen stage;  
(ii) \emph{Repair the rewrite} to handle the counterexamples while preserving the optimization; or  
(iii) \emph{Abandon the pair} if the counterexample reveals a fundamental semantic mismatch.

% This feedback loop follows the spirit of \emph{counterexample-guided synthesis} (CEGIS)~\cite{cegis}, where candidate solutions are iteratively challenged by counterexamples. Here, AdvCell supplies adversarial inputs that expose semantic mismatches, and the LLM responds by refining, constraining, or abandoning proposed rewrites. For efficiency, we limit this interaction to a single round of feedback in our implementation, though the design admits additional iterations if desired.

\paragraph{Final output.}
The output is a set of (original, rewritten code) pairs, $\pairs = \{(\cell, \cell'_i) \ | \ \cell'_i \in \snippetgenfunc(\cell) \wedge \phi_e(\cell'_i) \wedge \phi_p(\cell'_i)\}$. Figure~\ref{fig:examplePair} shows an example pair generated by \snippetgen that passes both EquivCheck and OptCheck constraints.

\subsection{Bridge: \rulegen}
\label{sec:rulegen}
So far, optimizations are tied to individual \pandas codes and cannot generalize to unseen, syntactically different code. \rulegen bridges the gap between LLM-driven rewrites and compiler-supported optimizations by extracting generalized rewrite rules from candidate pairs, enabling their application to semantically similar code patterns.

\paragraph{Problem Setup}
A rewrite rule $\rrule: \cells \rightarrow \cells$ maps an input cell to a rewritten cell. Formally, rules are expressed as $\alpha_\cell \Rightarrow_\phi \alpha_{\cell'}$, where $\alpha_\cell$ and $\alpha_{\cell'}$ are abstract representations of the original and rewritten cells, and $\phi$ is a precondition controlling rule applicability. \rulegen aims to learn a set $\rrules = \{\rrule_1, \dots, \rrule_N\}$ such that each $\rrule_i$ abstracts an optimization observed in $(\cell, \cell') \in \pairs$ and generalizes it. 
% \charith{ground this with in an example or put in the appendix. Again can use the example in the intro.}

% \charith{I don't see details about the checker part?? Isn't that important? I mean the adversarial agents? Also, cite if this design is prominent.}
\paragraph{AI Agent for Rule Generation.}
We implement rule generation using four sub-agents to transform a concrete pair $(\cell, \cell')$ into rewrite rules (Figure~\ref{fig:pipeline}).

\medskip
\noindent
\textbf{A1: Variable and Constant Generalizer.} This agent identifies variables $\vars$, constants $\constants$, and column identifiers $\identifiers$ in $\cell$ that can be abstracted, and maps them to corresponding elements in $\cell'$. This ensures rules remain correct under consistent renaming or generalization.

\noindent
\textbf{A2: AST Type Resolver.} It infers appropriate AST node types for abstracted variables using an LLM. This prevents over-specialization to the original cell while ensuring syntactic compatibility across similar code snippets.

\noindent
\textbf{A3: Rule Constructor.} This agent combines abstracted variables, constants, and identifiers with inferred types to create abstract representations $\alpha_\cell, \alpha_{\cell'} = \mathrm{abstract}(\cell, \cell', \vars, \constants, \identifiers)$. Syntactic correctness of the generated rule is verified using the compiler.

\noindent
\textbf{A4: Precondition Synthesizer.} It generates preconditions $\phi$ that encode syntactic and semantic constraints. $\phi$ is inferred to maximize rule validity across the original and synthetically perturbed program instances, with compiler checks ensuring syntactic correctness.

All agents include combined LLM and deterministic checks to validate outputs. If incorrect, it provides feedback to the original LLM to refine the output. Our implementation limits the maximum number of feedback iterations to 3. The prompts used for all the agents are provided in Appendix~\ref{appendix:prompts}.

\paragraph{Final Output.}
The stage produces a corpus of rewrite rules $\rrules = \{\rrule_1, \dots, \rrule_N\}$ suitable for automated application. Figure~\ref{fig:rewrite_rule} shows an example rewrite rule generated by \rulegen corresponding to the optimization in Figure~\ref{fig:examplePair}.

\subsection{Deployment Stage: \codegen}
\label{sec:codegen}
\begin{figure}[]
    \centering
    \begin{tcolorbox}[colback=backcolour, colframe=white, boxrule=0.1pt, arc=0.4mm, left=0.1mm, right=0.5mm, top=0.1mm, bottom=0.1mm]
        \begin{minted}[fontsize=\small,breaklines,breakanywhere, escapeinside=||]{python}
if isinstance(df, pd.DataFrame) and 'Date' in df.columns:
    df.pop('Date')
else:
    df = df.drop(['Date'], axis=1)
        \end{minted}
    \end{tcolorbox}

    \caption{Example application of a rewrite rule by \codegen.  
    % The original code snippet and the dataframe are taken from KGTorrent ~\cite{kgtorrent}.
    }
    \label{fig:exampleApplication}
\end{figure}

Once a set of rewrite rules has been generated and validated, \codegen applies them to unseen \python code snippets. Unlike earlier stages, this step requires \textbf{no LLM calls}, making it lightweight and deterministic. 
% In our implementation, we only retain the rewrite rules whose average speedup during OptCheck was $\geq 2$. \charith{say why a bit?}

\paragraph{Problem Setup}
Given a set of rewrite rules, \codegen produces an executable rewriter
\(\rewritecomp : \rrules \times \cells \rightarrow \cells\),
which, for any input cell \(\cell\), searches for applicable rules \(r \in \rrules\) whose patterns match \(\cell\) and returns a rewritten cell \(\rewritecomp(\cell)\).

\paragraph{Static Pattern Matching}
Each abstract pattern \(\alpha_{\cell}\) is compiled into a structural matcher \(\match(\cdot, \alpha_{\cell})\) that outputs:
\[
\match(\cell, \alpha_{\cell}) =
\begin{cases}
\theta & \text{if $\alpha_{\cell}$ matches $\cell$ under substitution } \theta, \\
\bot & \text{otherwise},
\end{cases}
\]
where \(\theta\) is a consistent binding of the abstract variables.
% (e.g., \(\alpha_{\text{col}} \mapsto \texttt{`price'}\) or \(\alpha_x \mapsto \mathtt{temp}\)).

\paragraph{Rule Application}
A rule \(r : \alpha_{\cell} \Rightarrow_\phi \alpha_{\cell'}\) is applied to a cell \(\cell\) under substitution \(\theta\) only if its precondition holds: \(\phi(\cell, \theta) = \texttt{true}\).  
Since \(\phi\) may depend on dynamic properties such as DataFrame shapes or column types, it cannot always be evaluated at compile time. \codegen therefore rewrites the cell into a conditional form: the optimized version is applied when \(\phi(\cell, \theta)\) holds, and the original cell is used otherwise. When the condition is satisfied at runtime, the optimized cell is instantiated as \(\theta(\alpha_{\cell'})\).
Figure~\ref{fig:exampleApplication} shows an example application of the rule shown in Figure~\ref{fig:rewrite_rule}.

\paragraph{Rule Scheduling}
Multiple rewrite rules may match a single input cell. \codegen employs a rule scheduler that determines which rules to apply. In our evaluation, we use a greedy scheduler that prioritizes rules based on their expected performance impact, as observed in the OptCheck during the \snippetgen stage. More sophisticated scheduling strategies can be incorporated in future work.
% To the best of our knowledge, \pandasbench is the only \pandas benchmark that consists of 102 real-world notebooks, whose main purpose is stress testing the \pandas optimization frameworks. None of the baselines can successfully run all the notebooks in this benchmark either because they do not handle the full \pandas functionality or they give errors on some notebooks.
\section{Evaluation}
\label{sec:evaluation}
% In this section, we present the evaluation to study the effectiveness of \framework. 
We address the following key questions: 
(1) What is the end-to-end performance of \framework on real-world benchmarks? (\S~\ref{sec:endtoend})
(2) How effective are the rules generated by \framework? (\S~\ref{sec:endtoend})
% (1) How effective are the rules generated by \framework? (\S~\ref{sec:endtoend})
(3) How frequently do the generated rules apply to real-world workloads? (\S~\ref{sec:hitrate})
(4) What is the yield of the \snippetgen and \rulegen stages? (\S~\ref{sec:yieldanalysis})
(5) What are typical successes and failure modes? (\S~\ref{sec:casestudies})
% 

% \aval{@Stefanos}
% \framework generated a total of 230 rewrite rules. For evaluation, we excluded rules that exhibited correctness issues identified during manual inspection (discussed in \S~\ref{sec:casestudies}), as well as rules with a reported OptCheck speedup of 2 or less, yielding a final set of 87 rules.

\paragraph{Datasets and Baselines.}
We collected 199 real-world \pandas notebooks from Kaggle using KGTorrent~\cite{kgtorrent}, which serve as the learning set for the \snippetgen and \rulegen stages. 

We evaluate \framework on Pandas notebooks from the \pandasbench benchmark~\cite{pandasbench}, which consists of Jupyter notebooks collected from Kaggle that use the \pandas API. 
To our knowledge, \pandasbench is the only benchmark composed of a large collection of real-world \pandas notebooks intended to evaluate \pandas optimization frameworks. Notably, none of the evaluated baselines can successfully execute all notebooks in \pandasbench, either due to unsupported \pandas functionality or runtime errors on certain workloads.
Further, while \pandasbench supports multiple scale factors that synthetically increase input sizes, we target the default (unscaled) setting to focus on realistic, unscaled workloads.

% For the final evaluation, we use \pandasbench, which consists of 102 notebooks with CSV-based inputs. \stef{There are two places discussing PandasBench extensively. There should be one.} 
% While \pandasbench supports multiple scale factors that synthetically increase input sizes, this work targets the default (unscaled) setting.
% This choice reflects our focus on realistic, unscaled workloads.
We compare \framework against all baselines considered in \pandasbench---\dias, \modin, \dask, and \koalas. Among these, \dias represents the prior compiler-based SOTA, while \modin is a systems-based SOTA approach.

The evaluation was executed on a dual-socket server with Intel Xeon Gold 6348 CPUs with 28 cores each, 1 TB of main memory, 6 TB of local NVMe storage (Dell PERC H755N Front), Ubuntu 22.04.5 LTS, and Python 3.10.19. 
% We use the default library versions from the \pandasbench repository. 
We used GPT-4.1 for the \snippetgen and \rulegen stages.
% \charith{people would like to see how many rules were generated first?}

% \aval{Mention that pandasbench is hard. We are the only one who get positive performance on pandasbench. Put pandasbench in background.}

\subsection{End-to-End Performance}
\label{sec:endtoend}
\begin{table}[t]
\centering
\caption{Performance comparison of \framework against baselines. The columns show the relative speedup achieved by \framework w.r.t the baselines. Out of 102, \framework executed 101.}
\label{tab:baselineNumbers}
\resizebox{\linewidth}{!}{
\begin{tabular}{@{}lrrrrr@{}} 
\toprule
\textbf{Framework} & \textbf{\# Notebooks} & \textbf{Mean} & \textbf{Median} & \textbf{Max} & \textbf{Min} \\
\midrule
\dias   & 97 & 1.54 & 1.13 & 4.3 & 0.80 \\
\modin  & 72 & 112.79 & 32.78 & 1914.89 & 0.48 \\
\dask   & 3  & 12.32 & 6.45 & 28.85 & 1.68 \\
\koalas & 10 & 140.15 & 100.36 & 377.09 & 7.25 \\
\bottomrule
\end{tabular}
}
\end{table}

% We begin by evaluating the end-to-end performance of \framework on the 102 notebooks in \pandasbench relative to the baselines.  
% For this evaluation, we used the rules that report a speedup $\ge 2$ in the discovery stage. \charith{can't you include this in the algo itself? Also, you can make it a parameter in algo.} 
% \charith{not clear to this audience why some baselines do not work} 
In Table~\ref{tab:baselineNumbers}, for each baseline, we report the number of notebooks successfully executed end-to-end and the speedup statistics of \framework over that baseline. \framework runs the maximum number of notebooks (101 out of 102) and achieves substantial improvements over all baselines. \framework achieves an mean speedup of $1.54\times$ over \dias, $112.79\times$ over \modin, $12.32\times$ over \dask, and $140.15\times$ over \koalas. These results establish \framework as SOTA optimization framework for real-world \pandas notebooks. 
% \stef{I wouldn't write such statements.}

\begin{figure}[t]
    \centering
    \includegraphics[width=0.8\linewidth]{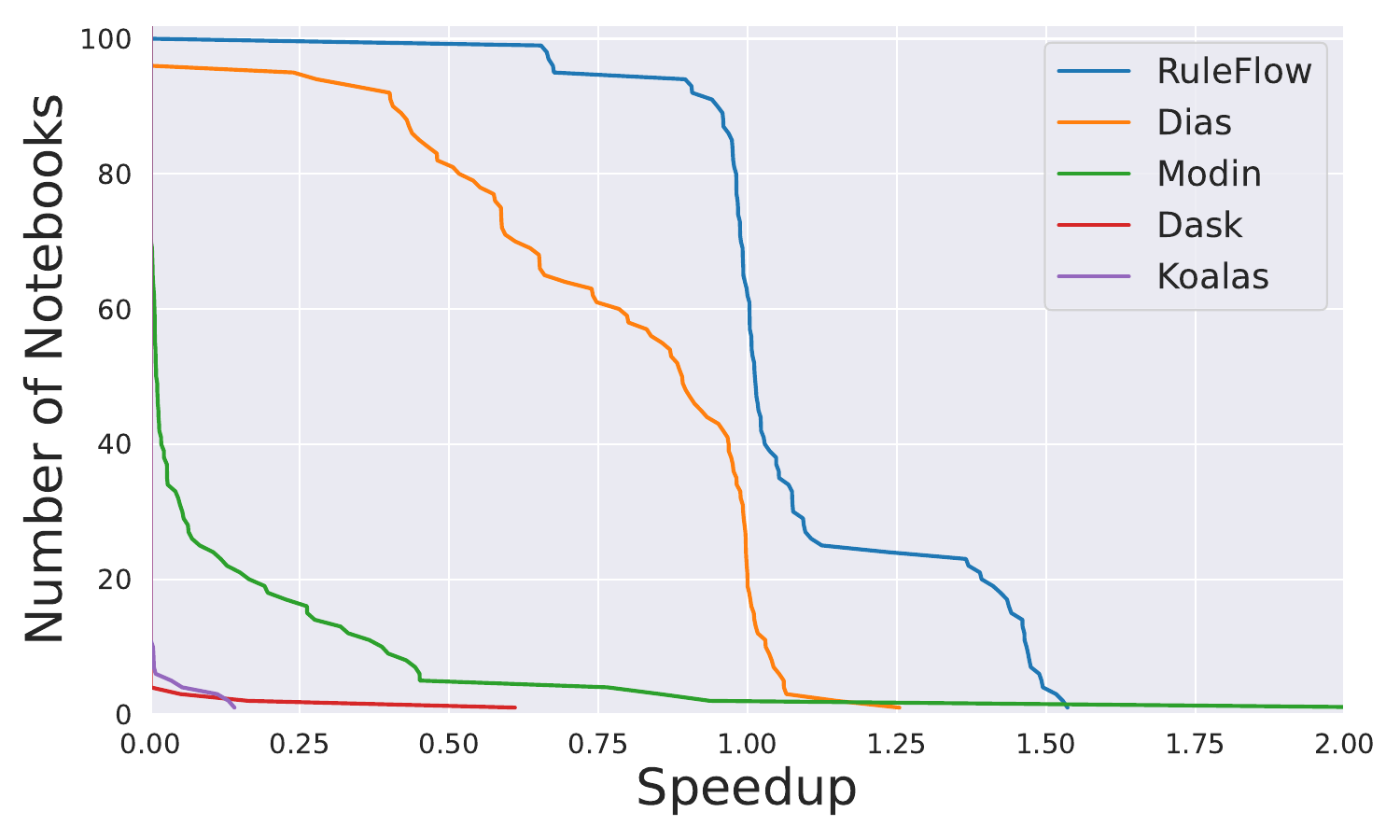}
    \caption{Speedups of different frameworks over \pandas across 102 notebooks. Higher curves indicate better overall performance. For the most part, \framework dominates existing frameworks, establishing a new SOTA \pandas optimization framework.}
    \label{fig:cactus}
\end{figure}

% \charith{can't we make this speedup over dias?}
Figure~\ref{fig:cactus} presents a cactus plot showing the distribution of end-to-end speedups across all 102 notebooks. For a given speedup threshold on the x-axis, the y-axis shows the number of notebooks achieving at least that speedup relative to \pandas. Curves that are higher on the plot indicate better overall performance. Overall, \framework outperforms existing baselines for most notebooks across the benchmark, reinforcing its SOTA performance.

\subsection{Rewrite Rule Analysis}
\begin{figure}[t]
    \centering
    \includegraphics[width=0.8\linewidth]{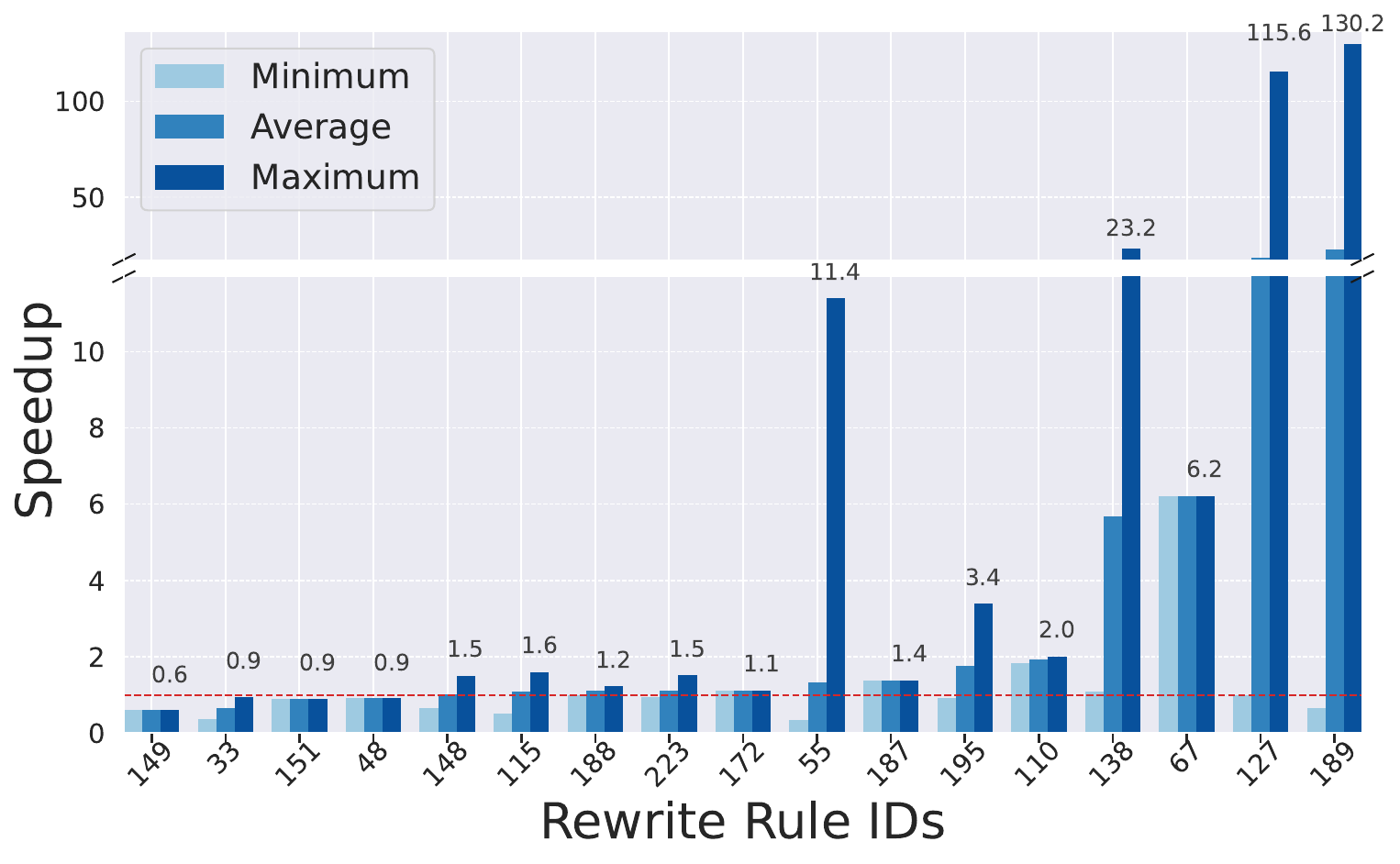}
    \caption{Per-rule speedups. Each bar group shows minimum, average, and maximum speedups across all applications of a rule, with the label indicating the maximum. Several rules achieve large speedups, with maximum speedups greater than $100\times$.}
    \label{fig:speedupPerRule}
\end{figure}
In this section, we present an in-depth analysis of the rewrite rules generated by \framework. First, we begin with aggregate statistics. \framework generated 120 rewrite rules in total. For evaluation, we excluded rules that exhibited correctness issues identified during qualitative analysis, yielding a final set of 88 rules. We discuss the manually detected errors and an in-depth qualitative study of the rules in \S~\ref{sec:casestudies}. We highlight that it is common for rewrite rules not to hit in certain workloads. We discuss this in detail in Appendix~\ref{appendix:lessRuleHits}.

Now we discuss how often these rules were applied. We say that when a rule applies to a piece of code, then it \textit{hits}. Amongst the $88$ rules, $24$ rules ($\sim 27\%$) hit at least one notebook.
Of these $24$, the \codegen scheduler scheduled $17$ rules (see \S\ref{sec:codegen}). We now concentrate on these 17 rules. Figure~\ref{fig:speedupPerRule} reports the minimum, average, and maximum speedup achieved by each rule across all cell-level applications in the dataset. Each bar group corresponds to a single rule, with the label indicating its maximum observed speedup.

Most rewrite rules provide consistent performance benefits, achieving average speedups greater than 1 across their applications. A single rule can provide up to $130\times$ speedup. These results illustrate that \framework identifies and exploits high-impact optimization opportunities across real-world notebooks, providing a clear explanation for the strong end-to-end performance observed in the benchmark. 

% Now we will dive deeper into how rules hit at the notebook level and then at the cell level.

% \stef{This should not be a random paragraph in end-to-end perf.}

% We also present a finer-grained analysis of individual rewrite rules.
% \aval{@Stefanos} \stef{I think the reader won't remember the number 88 you wrote many paragraphs ago. I would remove this paragraph from here and place in the next section.}
% Amongst the $88$ rules, $24$ rules were applied to atleast one notebook. Out of the $24$, the \codegen scheduler scheduled $17$ rules~\ref{sec:codegen}. 
% % \stef{I've no idea what the scheduler is. I think it's mentioned only here in the evaluation, so I'd point to the technical section that discusses it}. 
% Amongst all rules that hit, we show the speedups of the rules that were finally applied by the \codegen scheduler. 
% % \charith{baseline is pandas?} 
% Figure~\ref{fig:speedupPerRule} reports the minimum, average, and maximum speedup achieved by each rule across all cell-level applications in the dataset. Each bar group corresponds to a single rule, with the label indicating its maximum observed speedup.
% Most rewrite rules provide consistent performance benefits, achieving average speedups greater than 1 across their applications.
% Further, some rules exhibit large maximum speedups, approaching $130\times$ in some cases.
% These results illustrate that \framework identifies and exploits high-impact optimization opportunities across real-world notebooks, providing a clear explanation for the strong end-to-end performance observed in the benchmark.
\subsection{Hit Rate} 
\label{sec:hitrate}
% We show how often the rules apply to unseen notebooks.

\begin{figure}[t]
    \centering
    \includegraphics[width=0.8\linewidth]{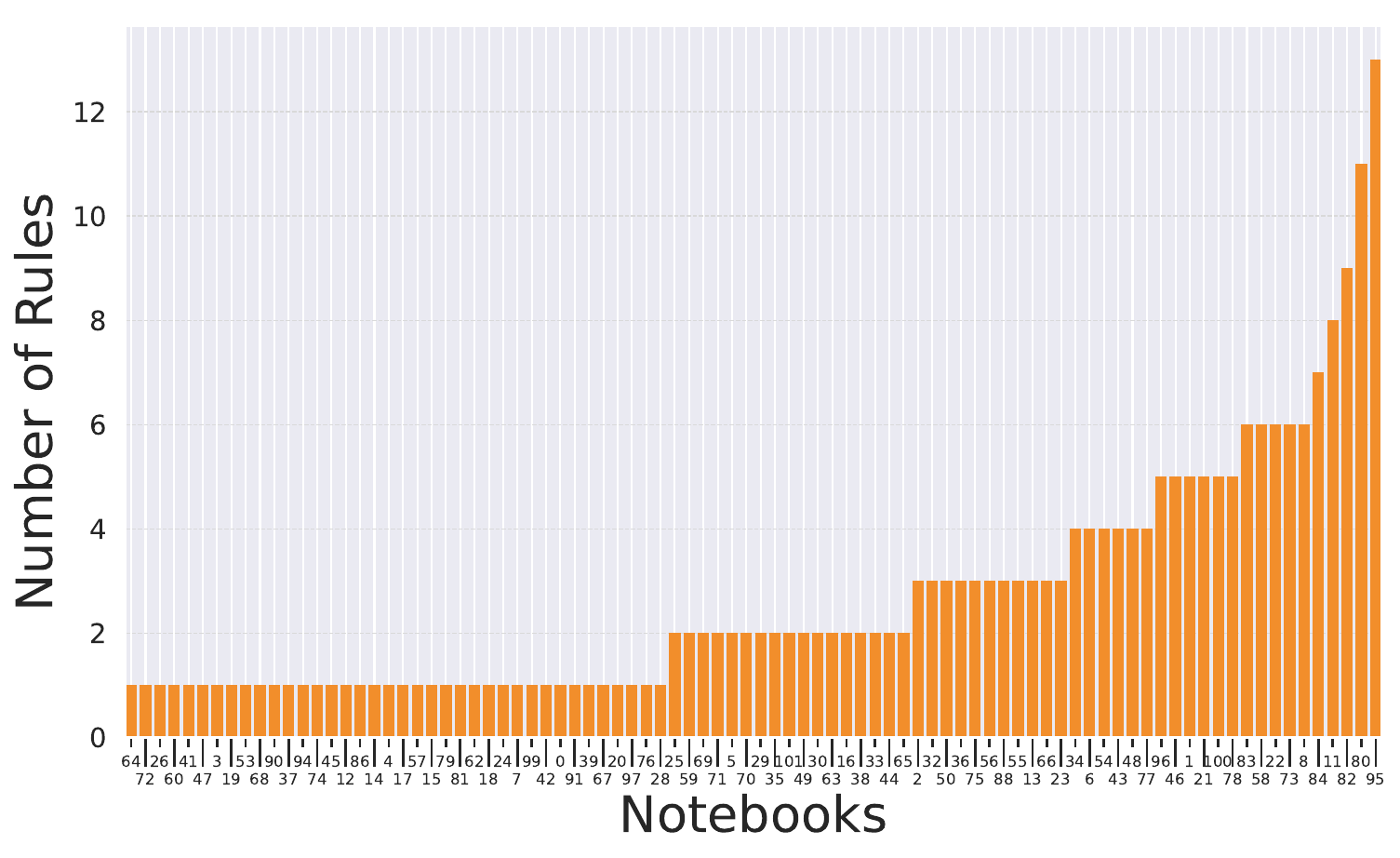}
    \caption{Notebook-wise rule hits showing the total number of rule applications in each \pandasbench notebook.}
    \label{fig:notebookHitRate}
\end{figure}

\paragraph{Notebook-Wise Hits.}
We measure the total number of rules that apply across all cells in a notebook. 
Multiple rules can apply to the same cell, and a single rule may fire in multiple cells; each application is counted separately. 
Figure~\ref{fig:notebookHitRate} shows how these counts are distributed across the 102 \pandasbench notebooks. Out of the 102 notebooks, 88 have at least one rule application, and in 17 notebooks, at least 5 different rules are applied. This suggests that the generated rules are often reused across real-world notebooks.

% \paragraph{Notebook-Wise Hits.}
% \stef{This is too obtuse. You can simply say sth like: "we say that rule hits on a notebook if it hits in any of its cells"} We first examine rule applicability at the notebook level, measured as the total number of rule applications across all cells in a notebook. Multiple rules can apply to the same cell, and a single rule may fire in multiple cells; each application is counted separately. Figure~\ref{fig:notebookHitRate} shows the distribution of hit rates across the 102 \pandasbench notebooks. 
% $88$ (out of $102$) notebooks have at least one rule application.
% A majority of the notebooks ($52$) see more than one rule application, \stef{this is weird, a notebook doesn't hit. You need sth like: in 17 notebooks at least 5 rules apply. And btw these should be rule hits right? in other words they are not distinct.} with $17$ notebooks hitting at least $5$ rules. This indicates that the generated rules are frequently reused in real-world workloads. 
% \charith{later section on qualitative analysis seems a bait-and-switch. Might need to explain that some rules do not hit.}
% \aval{Stefanos: please look at Charith's commment about case studies section} 
% \stef{you don't mention any later section here.}

\begin{figure}[t]
    \centering
    \includegraphics[width=0.8\linewidth]{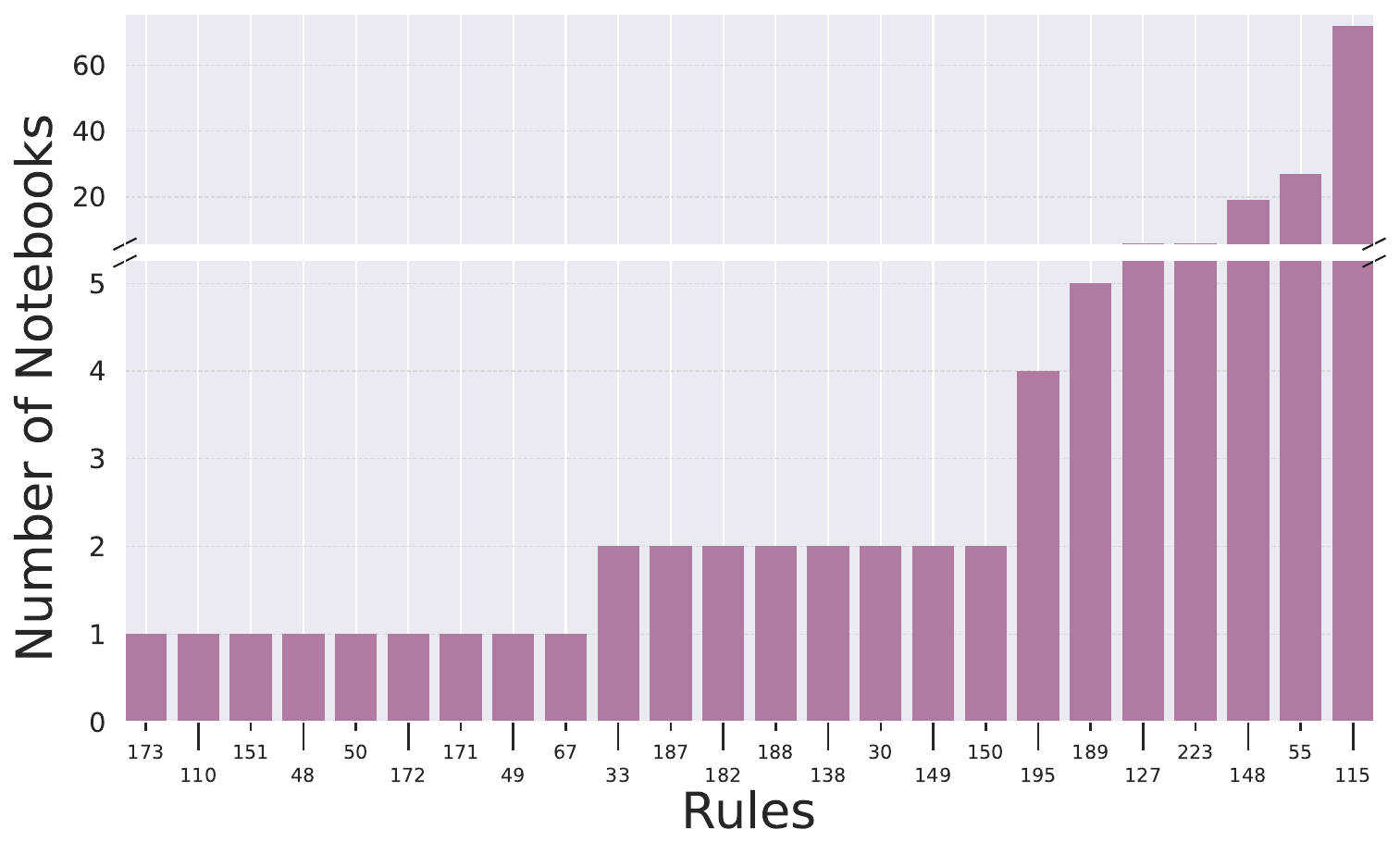}
    \caption{Rule-wise hit rates showing how often each rewrite rule is applied across all notebooks in \pandasbench.}
    \label{fig:ruleHitRate}
\end{figure}
\paragraph{Rule-Wise Hits.}
We also measure how often the rewrite rules apply across all notebooks. Figure~\ref{fig:ruleHitRate} shows that some rules generalize broadly and apply frequently, while others are more specialized, targeting specific patterns or workloads. A total of 7 rules apply to more than 5  notebooks, with 3 rules each applying to more than 10 notebooks. Along with the notebook-level hit rates, this analysis highlights the diverse utility of the generated rewrite rules, showing that the pipeline produces both widely applicable optimizations and rules tailored to real-world workloads.

{\footnotesize
\begin{table*}[]
\caption{Example rewrite rules and their speedups on \pandasbench.}
\begin{tabular}
{@{}|p{0.04\textwidth}|p{0.2\textwidth}|p{0.21\textwidth}|p{0.24\textwidth}|p{0.075\textwidth}|p{0.075\textwidth}|@{}}
\hline
%Header
\rowcolor{lightgray!15}
Rule & \centering LHS & \centering RHS & \centering Preconditions & \centering Avg Spd & \centering Max Spd \arraybackslash \\
\hline

%Row 1 - Patrician
\rowcolor{lightgray!15}
% \begin{minipage}[t]{\linewidth}
R1
% \end{minipage}
&
\begin{minipage}[t]{\linewidth}
\begin{minted}[fontsize=\scriptsize,breaklines,breakanywhere,bgcolor=lightgray!15, escapeinside=||]{python}
@{|\highlightast{Name}|: |\highlightvar{n1}|} = @{|\highlightast{Name}|: |\highlightvar{n1}|}.rename(columns=@{|\highlightast{expr}|: |\highlightvar{e1}|})
\end{minted}
\end{minipage}
&
\begin{minipage}[t]{\linewidth}
\begin{minted}
[fontsize=\scriptsize,breaklines,breakanywhere,bgcolor=lightgray!15, escapeinside=||]{python}
@{|\highlightvar{n1}|}.rename(columns=@{|\highlightvar{e1}|}, inplace=True)
\end{minted}
\end{minipage}
&
\begin{minipage}[t]{\linewidth}
\begin{minted}[fontsize=\scriptsize,breaklines,breakanywhere,bgcolor=lightgray!15, escapeinside=||]{python}
[is(@{|\highlightvar{n1}|}, pd.DataFrame), 
is(@{|\highlightvar{e1}|}, dict)]
\end{minted}
\end{minipage}
&
\begin{minipage}[t]{\linewidth}
\centering
$22.57\times$
\end{minipage}
&
\begin{minipage}[t]{\linewidth}
\centering
$130.22\times$
\end{minipage}
\\
\hline

%Row 2 - Plebeian
\rowcolor{lightgray!15}
\begin{minipage}[t]{\linewidth}
R2
\end{minipage}
&
\begin{minipage}[t]{\linewidth}
\begin{minted}[fontsize=\scriptsize,breaklines,breakanywhere,bgcolor=lightgray!15, escapeinside=||]{python}
@{|\highlightast{Name}|: |\highlightvar{n1}|} = @{|\highlightast{Name}|: |\highlightvar{n2}|}[@{|\highlightast{List}|: |\highlightvar{l1}|}]
\end{minted}
\end{minipage}
&
\begin{minipage}[t]{\linewidth}
\begin{minted}[fontsize=\scriptsize,breaklines,breakanywhere,bgcolor=lightgray!15, escapeinside=||]{python}
@{|\highlightvar{n1}|} = @{|\highlightvar{n2}|}.loc[:, @{|\highlightvar{l1}|}]
\end{minted}
\end{minipage}
&
\begin{minipage}[t]{\linewidth}
\begin{minted}[fontsize=\scriptsize,breaklines,breakanywhere,bgcolor=lightgray!15, escapeinside=||]{python}
[is(@{|\highlightvar{n2}|}, pd.DataFrame),
is(@{|\highlightvar{l1}|}, list),
all(label in @{|\highlightvar{n2}|}.columns for label in @{|\highlightvar{l1}|})]
\end{minted}
\end{minipage}
&
\begin{minipage}[t]{\linewidth}
\centering
$1.32\times$
\end{minipage}
&
\begin{minipage}[t]{\linewidth}
\centering
$11.40\times$
\end{minipage}
\\
\hline

\rowcolor{lightgray!15}
\begin{minipage}[t]{\linewidth}
R3
\end{minipage}
&
\begin{minipage}[t]{\linewidth}
\begin{minted}[fontsize=\scriptsize,breaklines,breakanywhere,bgcolor=lightgray!15, escapeinside=||]{python}
@{|\highlightast{Name}|: |\highlightvar{n1}|} = @{|\highlightast{Name}|: |\highlightvar{n2}|}[@{|\highlightast{expr}|: |\highlightvar{e1}|}][@{|\highlightast{Slice}|: |\highlightvar{s1}|}
\end{minted}
\end{minipage}
&
\begin{minipage}[t]{\linewidth}
\begin{minted}[fontsize=\scriptsize,breaklines,breakanywhere,bgcolor=lightgray!15, escapeinside=||]{python}
@{|\highlightvar{n1}|} = @{|\highlightvar{n2}|}.iloc[@{|\highlightvar{s1}|}, @{|\highlightvar{n2}|}.columns.get_indexer(@{|\highlightvar{e1}|})
\end{minted}
\end{minipage}
&
\begin{minipage}[t]{\linewidth}
\begin{minted}[fontsize=\scriptsize,breaklines,breakanywhere,bgcolor=lightgray!15, escapeinside=||]{python}
[isinstance(@{|\highlightvar{n2}|}, pd.DataFrame),
isinstance(@{|\highlightvar{e1}|}, list),
all(isinstance(col, str) for col in @{|\highlightvar{e1}|}),
all(col in @{|\highlightvar{n2}|}.columns for col in @{|\highlightvar{e1}|}),
@{|\highlightvar{n2}|}.columns.is_unique]
\end{minted}
\end{minipage}
&
\begin{minipage}[t]{\linewidth}
\centering
$0.60\times$
\end{minipage}
&
\begin{minipage}[t]{\linewidth}
\centering
$0.60\times$
\end{minipage}
\\
\hline
\end{tabular}
\label{tab:speedupRules}
\end{table*}
}
\subsection{Yield Analysis}
\label{sec:yieldanalysis}

% \begin{figure}[]
%     \centering
%     \begin{subfigure}[b]{\linewidth}
%         \centering
%         \includegraphics[width=\linewidth]{sections/figures/snippetgen_sankey_take_2.pdf}
%         \caption{Enter Caption}
%         \label{fig:sankeySnippetgen}
%     \end{subfigure}
%     % \hfill
%     \begin{subfigure}[b]{\linewidth}
%         \centering
%         \includegraphics[width=\linewidth]{sections/figures/rulegen_sankey_take_2.pdf}
%         \caption{Enter Caption}
%         \label{fig:sankeyRulegen}
%     \end{subfigure}
%     \caption{Sankey Charts}
%     \label{fig:sankey}
% \end{figure}

\begin{figure}[]
    \centering

    \begin{minipage}{\linewidth}
        \centering
        \includegraphics[width=\linewidth]{sections/figures/snippetgen_sankey_take_2.pdf}
        \caption{Yield analysis of \snippetgen}
        \label{fig:sankeySnippetgen}
    \end{minipage}

    \vspace{1em}

    \begin{minipage}{\linewidth}
        \centering
        \includegraphics[width=\linewidth]{sections/figures/rulegen_sankey_take_2.pdf}
        \caption{Yield analysis of \rulegen}
        \label{fig:sankeyRulegen}
    \end{minipage}

    % \caption{Sankey Charts}
    % \label{fig:sankey}
\end{figure}

% \begin{figure}[t]
%     \centering
%     \includegraphics[width=1\linewidth]{sections/figures/snippetgen_sankey_take_2.pdf}
%     \caption{Sankey chart for \snippetgen}
%     \label{fig:sankeySnippetgen}
% \end{figure}
% \begin{figure}[t]
%     \centering
%     \includegraphics[width=1\linewidth]{sections/figures/rulegen_sankey_take_2.pdf}
%     \caption{Sankey chart for \rulegen}
%     \label{fig:sankeyRulegen}
% \end{figure}

% We now analyze how candidate optimizations and rewrite rules progress through the \framework pipeline.   

\paragraph{\snippetgen.}
Figure~\ref{fig:sankeySnippetgen} illustrates the flow of candidate optimizations through the stages of \snippetgen. Starting from LLM-generated candidates, we track how many pass semantic equivalence checking (EquivCheck), how many yield measurable performance improvements (OptCheck), how many are rejected by the adversarial agent, and how many are ultimately retained as successful optimizations.

We begin with $199$ notebooks used for learning and extract cells that invoke the \pandas API. Each cell is scaled independently to ensure a runtime of at least 1 second; cells that do not meet this criterion are discarded. This process yields $1,237$ cells, which are passed to CandidateGen. For each cell, CandidateGen produces 1–5 candidate optimizations after deduplication.
Each original–candidate pair is then evaluated by EquivCheck and OptCheck. Pairs that pass the equivalence and optimization threshold are accepted. These pairs proceed through a feedback-driven refinement loop consisting of FeedbackGen followed by another round of CandidateGen, EquivCheck, and OptCheck. Across all iterations, $3,981$ pairs are rejected, and $157$ are accepted, corresponding to a final yield of $\sim 3.79\%$.

% \charith{highlight this!!!}
% The low yield reinforces that the per-program optimization approach is impractical because most LLM-generated candidates are either incorrect or fail to deliver speedups. 
% \textbf{The low yield reinforces that the per-program optimization approach is impractical, as most LLM-generated candidates are either functionally incorrect or fail to deliver speedups}.
\begin{tcolorbox}[colback=gray!5,colframe=gray!80, boxsep=1.5pt, left=3pt, right=3pt, top=3pt, bottom=3pt, arc=2pt, boxrule=0.5pt]
The low yield reinforces that the per-program optimization approach is impractical, as most LLM-generated candidates are either incorrect or fail to deliver speedups.
\end{tcolorbox}

\paragraph{\rulegen.}

\rulegen converts the accepted pairs from \snippetgen into rewrite rules. As described in \S~\ref{sec:rulegen}, this process is implemented using four agents. Starting from the $157$ accepted pairs, the first two agents identify generalizable elements, which are then used by the third and fourth agents to synthesize rewrite rules. Using our compiler, we discard syntactically incorrect rewrite rules. 
Overall, the pipeline converts $120$ optimization pairs into rewrite rules. We further did qualitative analysis to discard any incorrect rules, yielding $88$ rules in total (\S~\ref{sec:casestudies}). 
% \stef{Wait, so you discarded only 30 pairs? Then Figure 8b) is wrong.}. 
Combined with the per-rule speedups and hit rates, these results show that while per-program LLM-based optimization is inefficient, its benefits can be amplified by converting the optimizations into reusable rewrite rules that apply across unseen programs.

Although the four agents in \rulegen are conceptual and can be simulated by a single agent, we find empirically that explicitly decomposing the task yields higher-quality rules and better overall yield. We compare multi-agent and single-agent \rulegen configurations in Appendix~\ref{appendix:rulegen}.
\subsection{Qualitative Analysis}
\label{sec:casestudies}
% \aval{@Stefanos: This section may need rephrasing. }

% \aval{@Stefanos: Explicitly mention that it is fine to not hit all the rules. Cite LLVM, TASO.} \stef{I added a specific experiment.}

We present a qualitative analysis of representative rewrite rules generated by \framework to visualize the sources of the observed end-to-end speedups and to highlight the semantic and systems challenges that arise in automatically optimizing \pandas-based programs. 

% \aval{@Stefanos}
% In total, \framework discovered 230 candidate rewrite rules (Figure~\ref{fig:sankeyRulegen}). Of these, 87 rules achieved a speedup of at least $2\times$ during the discovery stage and did not give any error on manual inspection, were retained for further evaluation. However, only 24 of these rules applied at least once across the \pandasbench notebooks (Figure~\ref{fig:ruleHitRate}). We highlight that it is common for rewrite rules not to hit in certain workloads. For example, LLVM’s InstCombine has hundreds of rules, most of which hit rarely. As a concrete example, we checked the hit rate of a single rewrite rule in LLVM’s InstCombine pass.\footnote{This one: \texttt{~X + C --> (C-1) - X}~\cite{instcombine_rule} was chosen randomly.} We ran the InstCombine pass over the \emph{entire} sqlite3 code~\cite{sqlite3-source} (which has ~51K lines of code in C, translated to ~70K LLVM IR instructions; the rules are expressed over LLVM IR instructions). This rule hit \emph{zero} times. This was even after canonicalizing the code (SROA) to make it more amenable to optimization and even though the InstCombine algorithm is much more complex than the one we use.\footnote{It is an iterative algorithm, which means that if a rule does not hit in the initial version of the code, other rules may rewrite the code and allow it to hit on one of the future passes.}

\paragraph{Speedup Analysis.}
Table~\ref{tab:speedupRules} shows example rules that help explain the final performance gains.
A major source of speedups is the elimination of unnecessary data copying. For example, the rule R1 rewrites calls to the \texttt{rename} method to use the \texttt{inplace} attribute when applicable, which avoids copying a DataFrame. 
% The generated preconditions ensure that the transformation is only applied when the method arguments match expected types (e.g., a \pandas DataFrame and a \python dictionary). 
Such rules capture optimizations that are well known to expert users but are often overlooked in practice.
Another effective rule, R2, exploits specialized access patterns in \pandas. It replaces standard bracket-based column selection with \pandas-specific label-based \texttt{loc} accessor. While the resulting speedups are more modest, these rules tend to be broadly applicable.
% 

% \paragraph{Sources of Slowdowns.}
Not all discovered rules translate into performance gains during evaluation. 
There are cases where the rule applies, but the preconditions are not satisfied at runtime. This still leads to a slowdown due to the overhead of computing preconditions. 
% During discovery, the rule speedups correspond to a single code instance and do not include the runtime of preconditions. As a result, some rules may exhibit slowdowns in practice.
For instance, the rule R3 was applied to a cell; however, the preconditions were not satisfied, thereby giving an overall slowdown of $0.6\times$.
% For example, the rule R3 includes a guard that checks the uniqueness of the input DataFrame’s index, which incurs a worst-case time complexity of $O(N)$. The runtime analysis of the rule without preconditons is presented in Appendix~\ref{}. 
% Although the rewrite itself may be faster, the overhead of validating applicability can outweigh its benefit.

\paragraph{Invalid Rules.}
Since we used LLMs for generating rewrite rules, errors are expected. 
Beyond trivial errors, we observe two subtle sources of incorrectness. First, some rules assume the absence of \texttt{NaN} values. Since real-world datasets frequently contain missing values, such rules may produce outputs that differ from the original code. Second, certain rules are correct in isolation but can lead to unintended behavior in rare corner cases due to data-flow dependencies across notebook cells. 
We discuss them in Appendix~\ref{appendix:casestudies}.

Overall, the case studies demonstrate that while automatically discovered rewrite rules capture meaningful performance optimizations, LLM use occasionally leads to errors.
% 
% while also revealing the limitations of instance-based discovery and local reasoning.
This motivates future work on cost-aware precondition synthesis and richer semantic modeling.

\section{Conclusion}
We present \framework, a hybrid approach that combines the reliability of compilers with the creativity of LLMs. 
By separating LLM-based discovery from compiler-based deployment, \framework avoids the cost and unreliability of per-program LLM optimization. We introduce \rulegen, a novel bridge that distills the LLM-produced optimizations into reusable rewrite rules. Our evaluation on \pandasbench shows that \framework achieves SOTA performance, outperforming existing optimization techniques. 
% Together, these results demonstrate a practical and effective way to use the creativity of LLMs in scalable, reusable optimizers for \pandas workloads.

\clearpage
\section{Impact Statement}
This paper presents work whose goal is to advance the performance and reliability of the \pandas program optimization. The techniques introduced in this work focus on improving performance by combining compiler-based methods with LLMs.

As per the standard techniques in software engineering, we perform exhaustive testing of the optimizations proposed by the LLMs. However, providing formal correctness guarantees for the rewrite rules is out of the scope of this work due to the absence of formal semantics for \python and \pandas. Further, the rewrite rules are produced offline and integrated into a compiler-based pipeline, allowing them to be easily reviewed, tested, and validated using standard software engineering practices before use.

We do not anticipate significant negative societal or ethical consequences arising directly from this work. The primary effect of the proposed techniques is improved performance in EDA workloads. More broadly, this work explores a novel, scalable, and reliable way of using LLMs as optimizers.

\bibliography{ref}

\bibliographystyle{icml2026}

\newpage
\appendix
\onecolumn
% \section{Appendix}
% \label{sec:Appendix}

\section{Prompts}
\label{appendix:prompts}
This section provides prompts for the agents utilized in \framework.
\subsection{\snippetgen}

\begin{tcolorbox}[
    breakable,
    colback=gray!5,
    colframe=gray!75,
    title=Prompt for CandidateGen,
    arc=2mm,
    boxrule=0.5pt,
    top=2mm,
    bottom=2mm,
    left=2mm,
    right=2mm]

\begin{verbatim}
You are an automatic rewriter for Python code that uses the pandas and numpy
libraries. You need to help to rewrite the user's code to be more performant
while being semantically equivalent. Rewrite only the code that involves numpy,
pandas, or native Python. Do not forget anything that hampers the correctness of
your rewritten code.

You must respond in JSON format with two fields:
1. reasoning: A brief explanation of your optimization strategy and why the rewrite is faster
2. rewritten_snippet: The optimized code snippet, or False if you can't find a faster equivalent 
version

Do not make imports in the rewritten code. The rewritten code should be directly usable.
Try to retain the variables names in the original code to make our equivalence checking easier.
\end{verbatim}
\end{tcolorbox}

\begin{tcolorbox}[
    breakable,
    colback=gray!5,
    colframe=gray!75,
    title=Prompt for FeedbackGen,
    arc=2mm,
    boxrule=0.5pt,
    top=2mm,
    bottom=2mm,
    left=2mm,
    right=2mm]
\begin{verbatim}
You are an adversarial code analysis verifier. 
Your goal is to find counterexamples where optimization code would be INCORRECT.

Your task is to FIND COUNTEREXAMPLES where:
1. The before code would execute correctly
2. The optimized after code would produce DIFFERENT or INCORRECT results

Think adversarially about CORRECTNESS:
- What edge cases could break the transformation?
- What data types, shapes, or values would cause semantic differences?
- What pandas/numpy behaviors differ between the two approaches?
- Are there cases where the transformation changes behavior 
(ordering, indexing, dtypes, NaN handling, etc.)?

For each counterexample, provide:
1. A concrete description of the scenario
2. Why the LHS would work but RHS would fail or give different results
3. What change would prevent this counterexample

However, if the incorrect cases can just be handled by setting an appropriate 
precondition (small runtime), then consider the before and after case as equivalent.

Only produce output if at least one counterexample issue is found. 
Return your analysis in JSON format:
{{
    explanation: Your adversarial analysis explaining your search for counterexamples,
    is_valid: true if NO counterexamples found, false otherwise,
    counterexamples: [
        {{
            description: Description of the scenario,
             why_lhs_works: Why the LHS would work but RHS would fail or give different results,
            what_change_would_prevent_this_counterexample: What change would prevent this 
             counterexample
        }},
        
        ...
    ] (empty list [] if none found)
}}
\end{verbatim}

\end{tcolorbox}

\subsection{\rulegen}
\begin{tcolorbox}[
    breakable,
    colback=gray!5,
    colframe=gray!75,
    title=Prompt for Variable and Constant Generalizer,
    arc=2mm,
    boxrule=0.5pt,
    top=2mm,
    bottom=2mm,
    left=2mm,
    right=2mm]
\begin{verbatim}
You are an expert pandas code analyzer specializing in identifying variables that can be 
generalized in code transformation rules.

You are given two pandas code snippets. The before is the normal code and after is the optimized 
code. The overall goal is to find a generalized rule. The first step to do this is to find the
variables that can be generalized in the rule. 

Guidelines
1. Extract Concrete Expressions Only: Your answer must only contain the exact, literal substrings 
from the code. For example: df['A'], NA, or ['col1', 'col2'], x.
2. Do Not Invent Placeholders: Do not invent abstract names like `dataframe_expression` or 
`values_to_replace`. If you cannot point to the exact substring in the code, do not include it.
3. Capture Entire Expressions: Do not break down list expressions. If the code uses ['col_x', 
'col_y'], you must include the entire ['col_x', 'col_y'] string, not 'col_x' and 'col_y' separately.
4. Make sure to not include the formal arguments in your answer. It should only contain the 
actual arguments.
5. Normalize duplicates: Remove duplicates that differ only by quote style (`'col'` vs `col`). 
If a literal appears in both Before and After, keep it once.
6. Ignore method names, or transformations such as `.head()`, `.apply()`, or `.drop()`.
7. If a variable is mentioned in the after code but not in the before code, do not include it in 
your answer.
\end{verbatim}

\end{tcolorbox}

\begin{tcolorbox}[
    breakable,
    colback=gray!5,
    colframe=gray!75,
    title=Prompt for Checker of Variable and Constant Generalizer,
    arc=2mm,
    boxrule=0.5pt,
    top=2mm,
    bottom=2mm,
    left=2mm,
    right=2mm]
\begin{verbatim}
You are a code analysis verifier. Review the following analysis:

Original Code Transformation:
Before: {request.before_code}
After: {request.after_code}

Agent's Analysis:
Variables for generalization: {variables}
Explanation: {explanation}

Verification Steps

Part 1: Individual Variable Assessment
For each variable in the Variables for generalization list, verify the following:
1.  Existence: The variable exists as a literal substring in the 'before' and/or 'after' code.
2.  Concreteness: The variable is a concrete code expression (e.g., df['A'], NA, y) and not 
an abstract placeholder name (e.g., 'columns_expression', 'dataframe_name').
3.  Relevance: The variable is a relevant data expression or literal value, not an unrelated 
part of the code.
4.  Check if the variable is mentioned in the after code but not in the before code. If it is, 
provide feedback to remove it from the list.

Part 2: Overall List Assessment
After checking each variable, evaluate the list as a whole:
4.  Completeness: Does the list capture all the expressions or literal values required 
for generalization? Is anything missing?

Output Format
Return your verification in the following strict JSON format, following the pattern in the 
example entries:
{{  
variable_assessments: [
{{
    variable: df['comments'],
    is_valid: true,
    reasoning: This is a concrete data expression found in the 'before' and 'after' code.
}},
{{
    variable: dataframe_expression,
    is_valid: false,
    reasoning: This is an abstract placeholder, not a literal expression from the code.
}},
{{
    variable: 'TODO',
    is_valid: true,
    reasoning: This is a relevant string literal found in the code.
}}
],
is_list_complete: true/false,
completeness_reasoning: Based on criterion 4, explain if the list is complete or what is missing.,
overall_is_valid: true/false,
final_feedback: Provide a final summary. If invalid, state the main reason (e.g., 'Variable `df` 
is missing', '`dataframe_expression` is an invalid placeholder').
}}
\end{verbatim}

\end{tcolorbox}

\begin{tcolorbox}[
    breakable,
    colback=gray!5,
    colframe=gray!75,
    title=Prompt for AST Type Resolver,
    arc=2mm,
    boxrule=0.5pt,
    top=2mm,
    bottom=2mm,
    left=2mm,
    right=2mm]
\begin{verbatim}
You are an expert pandas code analyzer specializing in identifying variable types that can 
be generalized in code transformation rules.

You are given two pandas code snippets. The before is the normal code and after is the 
optimized code. The overall goal is to find a generalized rule. We have already figured out 
the variables in the snippet that can be generalized, your task to find out the python AST 
node type. Please make sure that the type is as general as possible. There are the types that 
are allowed: 
@{Name: x} => A Name denotes a Python identifier. It represents a variable, function, class, 
or module name that can appear in the code. Example: df, ser, item.
@{expr: e} => An expr denotes any valid Python expression, as defined by the Python grammar. 
It can be a variable, a literal, a function call, an attribute access, or a more complex expression. 
Example: df['col'], pandas.Series(a, ser.index).
@{Const(str): s} => A Const(str) denotes a string literal constant. Example: 'col', ':', 'a'.
@{Const(int): s} => A Const(int) denotes an intl constant. Example: 1, 2, 99.
@{Const(float): s} => A Const(float) denotes a float literal constant. Example: 2.3, 3.5.

Make sure to not include the formal arguments in your answer. It should only contain the 
actual arguments.
\end{verbatim}

\end{tcolorbox}

\begin{tcolorbox}[
    breakable,
    colback=gray!5,
    colframe=gray!75,
    title=Prompt for the Checker of AST Type Resolver,
    arc=2mm,
    boxrule=0.5pt,
    top=2mm,
    bottom=2mm,
    left=2mm,
    right=2mm]
\begin{verbatim}
You are a code analysis verifier. Review the following analysis:

Original Code Transformation:
Before: {request.before_code}
After: {request.after_code}

Variables for generalization: {variables_text}

Agent's Analysis:
AST node types: {ast_node_types}
Explanation: {explanation}

Verify if the identified AST node types are correct for generalization. Consider:
1. Are these the correct AST node types for the given variables that can be generalized?
2. Are the types as general as possible (Name, expr, Const(str), Const(int), Const(float))?
3. Do the AST node types match the variables provided (same count and order)?

Return your verification in JSON format:
{{
    explanation: Detailed explanation of your verification reasoning,
    is_valid: true/false,
    feedback: Specific feedback on what's correct or incorrect
}}
\end{verbatim}

\end{tcolorbox}

\begin{tcolorbox}[
    breakable,
    colback=gray!5,
    colframe=gray!75,
    title=Prompt for Rule Constructor,
    arc=2mm,
    boxrule=0.5pt,
    top=2mm,
    bottom=2mm,
    left=2mm,
    right=2mm]
\begin{verbatim}
You are an expert pandas code analyzer specializing in generating generalized transformation 
rules from code transformations, the variables in the snippet that can be generalized and the 
python AST node types of these variables.

You are given two pandas code snippets. The before is the normal code and after is the 
optimized code. The overall goal is to find a generalized rule. We have already figured out 
the variables in the snippet that can be generalized and the python AST node types of these variables 
Your job is to rewrite the LHS and RHS to a more generalized form. While renaming, make sure 
to use fresh variables and include their corresponding AST node types in LHS using the format: 
@{AST node type: fresh variable}. For the RHS, you use the same variables as the rewritten LHS. 
Here is the format: @{fresh variable}. Additionally, if a variable is mentioned more than once 
in the LHS, you will need to follow the format: @{AST node type: fresh variable}, each time 
it appears. You do not have the authority to change the variables and their python AST node 
types. Do not invent new generalized variables and their python AST node types. 
\end{verbatim}

\end{tcolorbox}

\begin{tcolorbox}[
    breakable,
    colback=gray!5,
    colframe=gray!75,
    title=Prompt for the Checker of Rule Constructor,
    arc=2mm,
    boxrule=0.5pt,
    top=2mm,
    bottom=2mm,
    left=2mm,
    right=2mm]
\begin{verbatim}
You are a code analysis verifier. Review the following analysis:

Original Code Transformation:
Before: {request.before_code}
After: {request.after_code}

Variables for generalization and their AST node types:
{variables_text}

Agent's Analysis:
Rewritten LHS: {rewritten_lhs}
Rewritten RHS: {rewritten_rhs}
Explanation: {explanation}

Verify if the generalized rule is correct. Consider:
1. Does the LHS correctly generalize the before code using @{{{{type: variable}}}} format?
2. Does the RHS correctly generalize the after code using @{{{{variable}}}} format?
3. Are the fresh variables used consistently between LHS and RHS?
4. Does the rule capture the essence of the transformation?
5. Does the LHS and RHS have the same variables and their python AST node types? If not, 
you need to provide feedback to change the LHS and RHS to have the same variables and their 
python AST node types.

Return your verification in JSON format:
{{
    explanation: Detailed explanation of your verification reasoning,
    is_valid: true/false,
    feedback: Specific feedback on what's correct or incorrect
}}
\end{verbatim}

\end{tcolorbox}

\begin{tcolorbox}[
    breakable,
    colback=gray!5,
    colframe=gray!75,
    title=Prompt for Precondition Synthesizer,
    arc=2mm,
    boxrule=0.5pt,
    top=2mm,
    bottom=2mm,
    left=2mm,
    right=2mm]
\begin{verbatim}
You are an expert pandas code analyzer specializing in generating runtime preconditions 
for code transformation rules.

You are given two pandas code snippets. The before is the normal code and after is the 
optimized code. The overall goal is to find a generalized rule. We have already figured out 
the variables in the snippet that can be generalized, the python AST node types of these variables 
and the generalized LHS and RHS. Your task is to define the runtime preconditions under which 
the rewritten LHS can always be transformed into the rewritten RHS. 

Guidelines:
1.runtime_preconditions is a non-empty list of Python boolean expressions that can each be 
used inside `assert (<expr>)` at runtime.
2.Use only names present in the rule (e.g., `@{e1}`, `@{df_var}`) plus standard, fully 
qualified pandas/numpy helpers: `pd`, `np`, `pandas`, `numpy`, `pd.api.types`.
3.Assume `import pandas as pd` and `import numpy as np` are already done by the caller. 
Do not include import statements in preconditions.
4.Pure expressions only: no `if`, `then`, `else`, `for`, `while`, `try`, `lambda`, f-strings, 
or assignments.
5.Do NOT use any prefixes like R: - just write pure Python boolean expressions.
6.Avoid redundant checks (e.g., if you check `type(x) == pd.Series`, don't also check 
`isinstance(x, pd.Series)`)
7.Prefer SIMPLER PRECONDITIONS over complex ones unless complexity is ABSOLUTELY NECESSARY.
8.Prefer using 'isinstance' over 'hasattr', wherever possibe.
9.Since the preconditions will always be evauated at runtime, make sure that the runtime of the 
preconditions is less than the actual LHS and RHS so that they don't induce a large overhead in 
runtime.
10.If there is no possible valid precondition with less runtime, just return False.
11.If you use external libraries in the preconditions, make sure that they are imported 
in the preconditions. 

Disallowed / common errors (must never appear):
1.Control-flow phrases: If … then …
2.English words in place of Python: All elements, and then, equals
3.Unqualified helpers: use .dtype, pd.api.types.is_numeric_dtype, not is_numeric_dtype
4.Preconditions about constants already fixed in the LHS (e.g., `axis == 1`, `inplace == True`)
5.Checks that don't ensure type semantics (e.g., only `hasattr`). 
6.Prefixes like R: - just write the boolean expression directly
7.No call-site introspection (no args/kwargs/*args/**kwargs).

Return your analysis in the following JSON format:
{{
    explanation: Brief explanation of the runtime preconditions,
    runtime_preconditions: [condition1, condition2, condition3]
    
}}
\end{verbatim}

\end{tcolorbox}

\begin{tcolorbox}[
    breakable,
    colback=gray!5,
    colframe=gray!75,
    title=Prompt for the Checker of Precondition Synthesizer,
    arc=2mm,
    boxrule=0.5pt,
    top=2mm,
    bottom=2mm,
    left=2mm,
    right=2mm]
\begin{verbatim}
You are an adversarial code analysis verifier. Your goal is to find counterexamples where the 
transformation rule would be INCORRECT.

Original Code Transformation:
Before (LHS): {request.before_code}
After (RHS - optimized): {request.after_code}

Variables for generalization and their AST node types:
{variables_text}

Rewritten LHS: {request.rewritten_LHS}
Rewritten RHS: {request.rewritten_RHS}

Agent's Analysis:
Runtime Preconditions: {runtime_preconditions}
Explanation: {explanation}

Your task is to FIND COUNTEREXAMPLES where:
1. The LHS (before code) would execute correctly
2. The RHS (optimized after code) would produce DIFFERENT or INCORRECT results
3. The current runtime preconditions do NOT prevent this case
4. The runtime preconditions would raise an error or not evaluate to a boolean

You should also evaluate if the preconditions are unnecessarily COMPLEX:
1. Are there redundant checks that could be removed?
2. Could complex boolean expressions be simplified?

Think adversarially about CORRECTNESS:
- What edge cases could break the transformation?
- What data types, shapes, or values would cause semantic differences?
- What pandas/numpy behaviors differ between the two approaches?
- Are there cases where the transformation changes behavior (ordering, indexing, dtypes, 
NaN handling, etc.)?

Think critically about SIMPLICITY:
- Could the transformation work safely with FEWER preconditions?
- Are any preconditions redundant or checking the same thing?
- Would removing a precondition actually allow incorrect transformations, or is it just defensive?

For each counterexample, provide:
1. A concrete description of the scenario (e.g., "When @{{e}} is a DataFrame with duplicate 
indices...")
2. Why the LHS would work but RHS would fail or give different results
3. What precondition would prevent this counterexample

For complexity issues:
1. Identify which precondition(s) are unnecessarily complex or redundant
2. Explain why they're not needed or how they could be simplified
3. Suggest the simpler alternative

Only produce output if at least one counterexample OR complexity issue is found. 
Return your analysis in JSON format:
{{
    "explanation": "Your adversarial analysis explaining your search for counterexamples AND 
    evaluation of complexity",
    "is_valid": true if NO counterexamples found AND preconditions are not overly complex, 
    false otherwise,
    "feedback": "Specific feedback: 
        - If counterexamples found: explain what additional preconditions are needed
        - If preconditions are too complex: explain which ones are unnecessary and suggest 
         simpler alternatives
           - If both valid and simple: explain why the preconditions are sufficient and appropriately 
            minimal
        - You can mark as invalid for EITHER correctness issues OR unnecessary complexity",
    "counterexamples": [
        "Correctness issue: When @{{e}} has duplicate indices, transformation fails",
           "Complexity issue: Precondition 3 'hasattr(@{{e}}, dtype) and type(@{{e}}) == pd.Series' 
           is redundant - the type check implies hasattr",
        ...
    ] (empty list [] if none found)
}}
\end{verbatim}

\end{tcolorbox}
\section{Less Frequent Rule Hits}
\label{appendix:lessRuleHits}
We highlight that it is common for rewrite rules not to hit in certain workloads. For example, LLVM’s InstCombine has hundreds of rules, most of which hit rarely. As a concrete example, we checked the hit rate of a single rewrite rule in LLVM’s InstCombine pass. This rule is: \texttt{~X + C --> (C-1) - X}~\cite{instcombine_rule} was chosen randomly. We ran the InstCombine pass over the \emph{entire} sqlite3 code~\cite{sqlite3-source} (which has ~51K lines of code in C, translated to ~70K LLVM IR instructions; the rules are expressed over LLVM IR instructions). This rule hit \emph{zero} times. This was even after canonicalizing the code (SROA) to make it more amenable to optimization and even though the InstCombine algorithm is much more complex than the one we use. It is an iterative algorithm, which means that if a rule does not hit in the initial version of the code, other rules may rewrite the code and allow it to hit on one of the future passes.
\section{Rulegen Ablation}
\label{appendix:rulegen}
In this section, we present a comparative study to analyse the benefits of using a multi-agent system instead of a single agent to generate rewrite rules from the proposed optimizations. 

We randomly sampled $50$ pairs generated by \snippetgen. Both \rulegen and a single agent were employed to synthesize rules from this subset. Subsequently, a qualitative study was performed on both sets of resulting rules. We find that splitting the task into 4 agents is substantially better than using a single agent. Only $18\%$ of the rules generated by a single agent were correct, whereas $68\%$ rules of the multi-agent system were correct. Below, we describe the common errors observed in both systems.

\subsection{Single Agent}
Our analysis revealed that  $18 \%$ of the rules synthesized by the agent were valid, with satisfactory grammatical correctness and sufficient generalization. 
Amongst the invalid outputs, the following errors were observed:

\paragraph{1a. Incorrect Abstract Syntax Tree (AST) Node Type Generation}
Oftentimes, the agent generated incorrect AST node types. This occured in two types. First, the agent suffered from the hallucination of node types, synthesizing fictitious types such as \texttt{attr} or \texttt{str} that are absent from Python’s \texttt{ast} module. Second, the agent committed naming violations by failing to adhere to strict naming conventions for valid AST nodes. These errors included the substitution of lowercase identifiers (e.g., \texttt{name}) for required capitalized forms (e.g., \texttt{Name}) and the omission of type specifications in constant nodes (e.g., generating \texttt{Const} rather than \texttt{Const(int)}). 
% \begin{figure}[H]
%     \centering
    \begin{tcolorbox}[colback=backcolour, colframe=white, boxrule=0.1pt, arc=0.4mm, left=0.1mm, right=0.5mm, top=0.1mm, bottom=0.1mm]
        \textbf{LHS:} 
        \begin{minted}[fontsize=\small,breaklines,breakanywhere,bgcolor=lightgray!15, escapeinside=||]{python}
@{name: lhs} = @{expr: df}[@{df}.@{attr: filter_col} == @{Const:
filter_val}][@{str: col}].unique().size
        \end{minted}
        \textbf{RHS:} 
        \begin{minted}[fontsize=\small,breaklines,breakanywhere,bgcolor=lightgray!15, escapeinside=||]{python}
@{lhs} = len(set(@{df}.loc[@{df}.@{filter_col} == @{filter_val},
@{col}]))
        \end{minted}

        \textbf{Runtime Preconditions:}
        \begin{minted}[fontsize=\small,breaklines,breakanywhere,bgcolor=lightgray!15, escapeinside=||]{python}
isinstance(@{df}, pd.DataFrame), 
@{filter_col} in @{df}.columns, @{col} in @{df}.columns
        \end{minted}
    \end{tcolorbox}

%     \caption{}
%     \label{fig:Abalation1a}
% \end{figure}

As illustrated above, the Left-Hand Side (LHS) invalidates the rule by hallucinating the AST Node types (\texttt{attr, str}), violating capitalization constraints (using \texttt{name} instead of \texttt{Name}), and omitting the type for the constant node (using \texttt{Const} instead of \texttt{Const(int)}).

\paragraph{1b. Incorrect Rule Generation} The agent failed to comply with the DSL grammar. More specifically, according to the grammar, any variable appearing multiple times within the LHS must be abstracted using the format \texttt{@\{AST-node-type: variable\}}. Despite being explicitly stated in the prompt, the agent failed to adhere to it. As illustrated below, the agent correctly abstracted the first occurrence of the variable \texttt{df} in the LHS but mistakenly used the identifier \texttt{@\{df\}} for subsequent mentions in the LHS.

% \begin{figure}[H]
%     \centering
    \begin{tcolorbox}[colback=backcolour, colframe=white, boxrule=0.1pt, arc=0.4mm, left=0.1mm, right=0.5mm, top=0.1mm, bottom=0.1mm]
        \textbf{LHS:} 
        \begin{minted}[fontsize=\small,breaklines,breakanywhere,bgcolor=lightgray!15, escapeinside=||]{python}
@{Name: columns_drop} = @{Name: df}.dropna(axis=1)
@{Name: var2} = 'Columns in original dataset: %d\n' % @{df}.shape[1]
@{Name: var3} = "Columns with na's dropped: %d" % @{columns_drop}.shape[1]
        \end{minted}
        \textbf{RHS:} 
        \begin{minted}[fontsize=\small,breaklines,breakanywhere,bgcolor=lightgray!15, escapeinside=||]{python}
@{Name: n_cols_original} = @{df}.shape[1]
@{Name: n_cols_no_na} = (@{df}.isna().sum() == 0).sum()
@{var2} = 'Columns in original dataset: %d\n' % @{n_cols_original}
@{var3} = "Columns with na's dropped: %d" % @{n_cols_no_na}
        \end{minted}

        \textbf{Runtime Preconditions:}
        \begin{minted}[fontsize=\small,breaklines,breakanywhere,bgcolor=lightgray!15, escapeinside=||]{python}
isinstance(@{df}, pd.DataFrame),
len(@{df}.shape) > 1
        \end{minted}
    \end{tcolorbox}

%     \caption{}
%     \label{fig:Abalation1b}
% \end{figure}

\paragraph{1c. Incorrect Precondition Generation} The agent utilized the \texttt{hasattr} function despite explicit prompt-level instructions to the contrary. This reliance on verifying the existence of an attribute rather than the object's class can lead to incorrectness. As shown below, the precondition \texttt{hasattr(@{df}...plot, 'bar')} is insufficient, as it allows the rule to trigger for any object that has a \texttt{bar} attribute, regardless of whether that object supports the targeted transformation.
Further, the agent occasionally synthesized preconditions that were computationally expensive, making the rule impractical. 
Below, we show an example where validating the precondition would require executing the underlying dataframe operations (\texttt{groupby} and \texttt{sum}), which introduces significant latency. 
% \begin{figure}[t]
%     \centering
    \begin{tcolorbox}[colback=backcolour, colframe=white, boxrule=0.1pt, arc=0.4mm, left=0.1mm, right=0.5mm, top=0.1mm, bottom=0.1mm]
        \textbf{LHS:} 
        \begin{minted}[fontsize=\small,breaklines,breakanywhere,bgcolor=lightgray!15, escapeinside=||]{python}
@{name: lhs} = @{expr: df}.groupby(@{expr: by})[@{expr: col}].sum().plot(kind='bar')

        \end{minted}
        \textbf{RHS:} 
        \begin{minted}[fontsize=\small,breaklines,breakanywhere,bgcolor=lightgray!15, escapeinside=||]{python}
@{lhs} = @{df}.groupby(@{by})[@{col}].sum().plot.bar()
        \end{minted}

        \textbf{Runtime Preconditions:}
        \begin{minted}[fontsize=\small,breaklines,breakanywhere,bgcolor=lightgray!15, escapeinside=||]{python}
hasattr(@{df}.groupby(@{by})[@{col}].sum().plot, 'bar')
        \end{minted}
    \end{tcolorbox}
%     \caption{}
%     \label{fig:Abalation1c}
% \end{figure}

\paragraph{1d. Incomplete Variable and Constant Generalization} Another category of failures is the inability of the agent to abstract all generalizable variables and constants. The agent occasionally left hard-coded specific values (such as slice indices) into the rule rather than abstracting them. This results in "over-fitted" rules that cannot be applied broadly.
% \begin{figure}[b]
%     \centering
    \begin{tcolorbox}[colback=backcolour, colframe=white, boxrule=0.1pt, arc=0.4mm, left=0.1mm, right=0.5mm, top=0.1mm, bottom=0.1mm]
        \textbf{LHS:} 
        \begin{minted}[fontsize=\small,breaklines,breakanywhere,bgcolor=lightgray!15, escapeinside=||]{python}
@{Name: missing_values_count} = @{Name: df}.isnull().sum()
@{Name: out_var} = @{missing_values_count}[0:10]

        \end{minted}
        \textbf{RHS:} 
        \begin{minted}[fontsize=\small,breaklines,breakanywhere,bgcolor=lightgray!15, escapeinside=||]{python}
@{out_var} = @{df}.loc[:, @{df}.columns[0:10]].isnull().sum()
        \end{minted}

        \textbf{Runtime Preconditions:}
        \begin{minted}[fontsize=\small,breaklines,breakanywhere,bgcolor=lightgray!15, escapeinside=||]{python}
isinstance(@{df}, pd.DataFrame), 
@{df}.shape[1] >= 10
        \end{minted}
    \end{tcolorbox}
%     \caption{}
%     \label{fig:Abalation1d}
% \end{figure}
As shown above, the agent failed to abstract the Slice \texttt{0:10}, hampering the generalizability of the rule. 

\subsection{Rulegen} 
\begin{figure}[t]
    \centering
    \includegraphics[width=\textwidth]{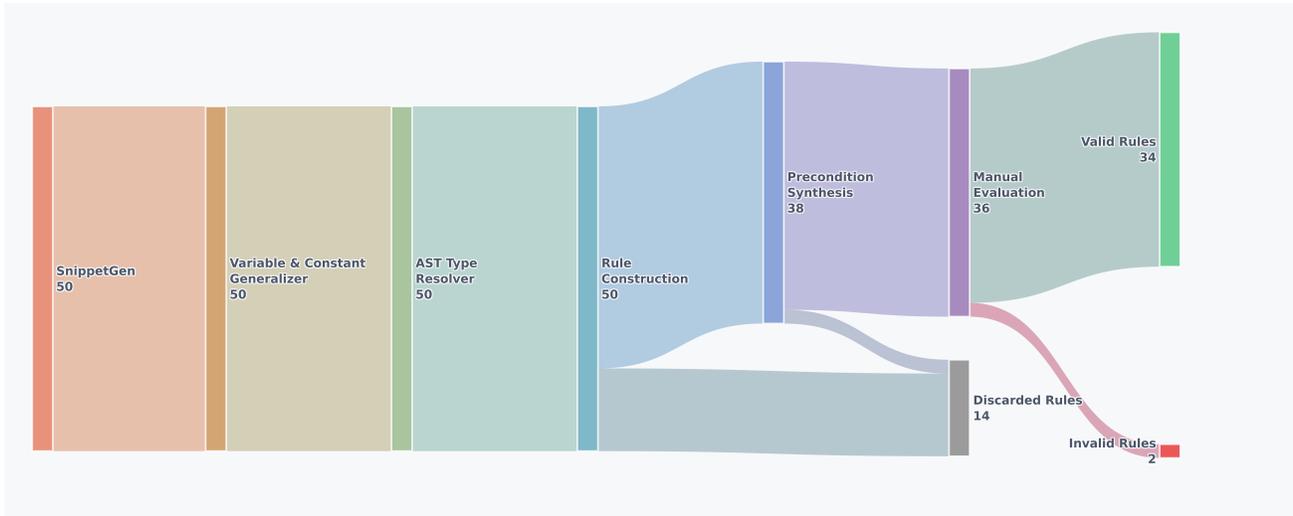}
    \caption{Sankey diagram illustrating the filtration process within \rulegen. The flow tracks the 50 initial pairs through \rulegen, followed by a final manual evaluation.}
    \label{fig:rulegen_flow}
\end{figure}

Of the initial $50$ candidate pairs, \rulegen's internal validity checks eliminated 14. As illustrated in Figure \ref{fig:rulegen_flow}, the Rule Construction Agent (A3) was responsible for the majority of these rejections $(n=12)$, while the Precondition Synthesis Agent (A4) discarded the remaining 2. 
Subsequent a qualitative analysis of the remaining 36 candidates identified two invalid rules. The final yield consisted of $34$ valid rules out of the original $50$, resulting in an overall validity rate of $68\%$. Following are the errors observed in the generated rules.

\paragraph{2a. Incorrect Precondition Generation} Like the single agent, \rulegen ignored precondition constraints by using attribute checks and synthesizing expensive preconditions. As shown below, the agent uses \texttt{hasattr} checks in addition to making the preconditions expensive.

% \begin{figure}[H]
%     \centering
    \begin{tcolorbox}[colback=backcolour, colframe=white, boxrule=0.1pt, arc=0.4mm, left=0.1mm, right=0.5mm, top=0.1mm, bottom=0.1mm]
        \textbf{LHS:} 
        \begin{minted}[fontsize=\small,breaklines,breakanywhere,bgcolor=lightgray!15, escapeinside=||]{python}
@{Name: n1} = @{Name: n2}.groupby(@{Const(str): c1})[@{Const(str): c2}].sum()
.plot(kind=@{Const(str): c3})

        \end{minted}
        \textbf{RHS:} 
        \begin{minted}[fontsize=\small,breaklines,breakanywhere,bgcolor=lightgray!15, escapeinside=||]{python}
@{n2}.groupby(@{c1})[@{c2}].sum().plot.@{c3}()
        \end{minted}

        \textbf{Runtime Preconditions:}
        \begin{minted}[fontsize=\small,breaklines,breakanywhere,bgcolor=lightgray!15, escapeinside=||]{python}
hasattr(@{n2}.groupby(@{c1})[@{c2}].sum().plot, @{c3}),
callable(getattr(@{n2}.groupby(@{c1})[@{c2}].sum().plot, @{c3}))
        \end{minted}
    \end{tcolorbox}
%     \caption{}
%     \label{fig:Abalation2a}
% \end{figure}

\paragraph{2b. Incomplete Variable and Constant Generalization} In some cases, the Variable and Constant Generalizer Agent (A1) failed to fully abstract all generalizable variables and constants. For example, as shown below, specific column names (\texttt{'date\_crawled'}, etc.) were hard-coded into the RHS and preconditions rather than being abstracted into variables, resulting in an "over-fitted" rule.

% \begin{figure}[H]
%     \centering
    \begin{tcolorbox}[colback=backcolour, colframe=white, boxrule=0.1pt, arc=0.4mm, left=0.1mm, right=0.5mm, top=0.1mm, bottom=0.1mm]
        \textbf{LHS:} 
        \begin{minted}[fontsize=\small,breaklines,breakanywhere,bgcolor=lightgray!15, escapeinside=||]{python}
@{Name: n1} = @{Subscript: s1}[@{Slice: s2}]

        \end{minted}
        \textbf{RHS:} 
        \begin{minted}[fontsize=\small,breaklines,breakanywhere,bgcolor=lightgray!15, escapeinside=||]{python}
@{n1} = @{s1}.iloc[@{s2}, @{s1}.columns.get_indexer(['date_crawled', 
'ad_created', 'last_seen'])],
        \end{minted}

        \textbf{Runtime Preconditions:}
        \begin{minted}[fontsize=\small,breaklines,breakanywhere,bgcolor=lightgray!15, escapeinside=||]{python}
isinstance(@{s1}, pd.DataFrame), 
isinstance(['date_crawled', 'ad_created', 'last_seen'], list), 
all(isinstance(col, str) for col in ['date_crawled', 'ad_created', 'last_seen']), 
all(col in @{s1}.columns for col in ['date_crawled', 'ad_created', 'last_seen'])
        \end{minted}
    \end{tcolorbox}
%     \caption{}
%     \label{fig:Abalation2b}
% \end{figure}

\subsection{Comparative Analysis}
A direct comparison highlights the efficacy of the multi-agent architecture. The baseline Single Agent achieved a validity rate of $18\%$. In contrast, \rulegen's multi-agent architecture achieves a $68\%$ yield. Notably, \rulegen eliminated the syntactic Rule Generation and AST-conformance errors (Types 1a and 1b) that the Single Agent suffered from. Furthermore, \rulegen's decomposition into different agents increases transparency by breaking rule synthesis into more interpretable steps. 

\subsection{Prompt for Single Agent}
\begin{tcolorbox}[
    breakable,
    colback=gray!5,
    colframe=gray!75,
    title=Prompt for Single Agent,
    arc=2mm,
    boxrule=0.5pt,
    top=2mm,
    bottom=2mm,
    left=2mm,
    right=2mm]
\begin{verbatim}
You are given two versions of code: Before and After.

You are an expert pandas code analyzer specializing in generating generalized transformation rules 
from code transformations.
You are given two pandas code snippets. The "before" is the original code and "after" is the 
optimized/transformed code. Your task is to generate a complete generalized rule. 
Each rule contains three parts, rewritten_LHS, rewritten_RHS, and runtime_preconditions. 

Guidelines for rewritten_LHS and rewritten_RHS:
1. The generalized variables in the LHS : @{AST node type: fresh variable}. 
2. The generalized variables in the RHS are the same as the generalized variables in the LHS 
but without the @{AST node type: } prefix.
3. If a variable is mentioned more than once in the LHS, you will need to follow the format: 
@{AST node type: fresh variable}, each time it appears. 

Guidelines for runtime preconditions:
1.Runtime_preconditions is a non-empty list of Python boolean expressions that can each be used 
inside `assert (<expr>)` at runtime.
2.Use only names present in the rule (e.g., `@{e1}`, `@{df_var}`) plus standard, fully qualified 
pandas/numpy helpers: `pd`, `np`, `pandas`, `numpy`, `pd.api.types`.
3.Assume `import pandas as pd` and `import numpy as np` are already done by the caller. Do not 
include import statements in preconditions.
4.Pure expressions only: no `if`, `then`, `else`, `for`, `while`, `try`, `lambda`, f-strings, or 
assignments.
5.Do NOT use any prefixes like R: - just write pure Python boolean expressions.
6.Avoid redundant checks (e.g., if you check `type(x) == pd.Series`, don't also check 
`isinstance(x, pd.Series)`)
7.Prefer SIMPLER PRECONDITIONS over complex ones unless complexity is ABSOLUTELY NECESSARY.
8.Prefer using 'isinstance' over 'hasattr', wherever possibe.
9.Since the preconditions will always be evauated at runtime, make sure that the runtime of the
preconditions is less than the actual LHS and RHS so that they don't induce a large overhead in 
runtime.
10.If there is no possible valid precondition with less runtime, just return False.

Disallowed / common errors (must never appear) for preconditions:
1.Control-flow phrases: "If … then …"
2.English words in place of Python: "All elements", "and then", "equals"
3.Unqualified helpers: use .dtype, pd.api.types.is_numeric_dtype, not is_numeric_dtype
4.Preconditions about constants already fixed in the LHS (e.g., `axis == 1`, `inplace == True`)
5.Checks that don't ensure type semantics (e.g., only `hasattr`). 
6.Prefixes like "R:" - just write the boolean expression directly
7.No call-site introspection (no args/kwargs/*args/**kwargs).
\end{verbatim}

\end{tcolorbox}

\section{Qualitative Analysis}
\label{appendix:casestudies}

In this section, we provide more details regarding the types of semantic failures that occur due to our rules. As expected, the usage of LLMs for rule synthesis introduced some errors. These are described below.

\paragraph{Incorrect Generalization} In some rules, the LLM incorrectly generalises the rule, resuling into an incorrect RHS. Since these missing generalisations arise from the original code during the discovery stage, they do not give errors during EquivCheck. However, for every other application of the rule, the RHS will either crash or give an incorrect output. An example of such a rule is shown below.
\begin{tcolorbox}[colback=backcolour, colframe=white, boxrule=0.1pt, arc=0.4mm, left=0.1mm, right=0.5mm, top=0.1mm, bottom=0.1mm]
    \textbf{LHS:} 
    \begin{minted}[fontsize=\small,breaklines,breakanywhere,bgcolor=lightgray!15, escapeinside=||]{python}
@{|\highlightast{Name}|: |\highlightvar{n1}|} = @{|\highlightast{Name}|: |\highlightvar{n2}|}.groupby(@{|\highlightast{expr}|: |\highlightvar{e1}|}).agg(@{|\highlightast{expr}|: |\highlightvar{e2}|}).reset_index()

    \end{minted}
    \textbf{RHS:} 
    \begin{minted}[fontsize=\small,breaklines,breakanywhere,bgcolor=lightgray!15, escapeinside=||]{python}
@{|\highlightvar{n1}|} = @{|\highlightvar{n2}|}.groupby(@{|\highlightvar{e1}|}).agg({'match_id': 'nunique', 'batsman_runs': 'sum', 'Fours': 'sum', 'Sixes': 'sum'}).reset_index()
    \end{minted}

    \textbf{Runtime Preconditions:}
    \begin{minted}[fontsize=\small,breaklines,breakanywhere,bgcolor=lightgray!15, escapeinside=||]{python}
'match_id' in @{|\highlightvar{e2}|},
callable(@{|\highlightvar{e2}|}['match_id']),
@{|\highlightvar{e2}|}['match_id'].__name__ == (lambda x: x.nunique()).__name__,
all(k in @{|\highlightvar{e2}|} for k in ['batsman_runs','Fours','Sixes']),
@{|\highlightvar{e2}|}['batsman_runs'] == 'sum',
@{|\highlightvar{e2}|}['Fours'] == 'sum',
@{|\highlightvar{e2}|}['Sixes'] == 'sum'
    \end{minted}
\end{tcolorbox}

\paragraph{LHS same as RHS}. In some rules, the RHS generated by the pipeline was the same as the LHS, i.e., the original code. These rules will always give a minor slowdown due to the added runtime of computing the preconditions. An example of such a rule is shown below.
\begin{tcolorbox}[colback=backcolour, colframe=white, boxrule=0.1pt, arc=0.4mm, left=0.1mm, right=0.5mm, top=0.1mm, bottom=0.1mm]
    \textbf{LHS:} 
    \begin{minted}[fontsize=\small,breaklines,breakanywhere,bgcolor=lightgray!15, escapeinside=||]{python}
@{|\highlightast{Name}|: |\highlightvar{d\_var\_1}|} = @{|\highlightast{Name}|: |\highlightvar{df}|}.isnull().values.any()

    \end{minted}
    \textbf{RHS:} 
    \begin{minted}[fontsize=\small,breaklines,breakanywhere,bgcolor=lightgray!15, escapeinside=||]{python}
@{|\highlightvar{d\_var\_1}|} = @{|\highlightvar{df}|}.isnull().any().any()
    \end{minted}

    \textbf{Runtime Preconditions:}
    \begin{minted}[fontsize=\small,breaklines,breakanywhere,bgcolor=lightgray!15, escapeinside=||]{python}
isinstance(@{|\highlightvar{df}|}, pandas.DataFrame)
    \end{minted}
\end{tcolorbox}

\paragraph{Trivial Precondition}. In some rules, the pipeline generated a trivial precondition, where the precondition ensures semantic equality between the LHS and the RHS by explicitly evaluating both at runtime, thereby increasing computation overhead and rendering the rule impractical. An example of such a rule is shown below.
\begin{tcolorbox}[colback=backcolour, colframe=white, boxrule=0.1pt, arc=0.4mm, left=0.1mm, right=0.5mm, top=0.1mm, bottom=0.1mm]
    \textbf{LHS:} 
    \begin{minted}[fontsize=\small,breaklines,breakanywhere,bgcolor=lightgray!15, escapeinside=||]{python}
@{|\highlightast{Name}|: |\highlightvar{v1}|} = @{|\highlightast{Name}|: |\highlightvar{v2}|}.shape[0]

    \end{minted}
    \textbf{RHS:} 
    \begin{minted}[fontsize=\small,breaklines,breakanywhere,bgcolor=lightgray!15, escapeinside=||]{python}
@{|\highlightvar{v1}|} = @{|\highlightvar{v2}|}.__len__()
    \end{minted}

    \textbf{Runtime Preconditions:}
    \begin{minted}[fontsize=\small,breaklines,breakanywhere,bgcolor=lightgray!15, escapeinside=||]{python}
hasattr(@{|\highlightvar{v2}|}, 'shape'),
isinstance(@{|\highlightvar{v2}|}.shape, tuple),
@{|\highlightvar{v2}|}.shape[0] == @{|\highlightvar{v2}|}.__len__()
    \end{minted}
\end{tcolorbox}

% \paragraph{1. Trivial Errors} As expected, the usage of LLMs for rule synthesis introduced certain trivial errors. For instance, Figure~\ref{} shows a no-op rule where the LHS is syntactically equivivalent to the RHS. . 
% % \begin{minted}{python} 
% % LHS: @{Name: v1} = @{expr: e1}.nunique()
% % RHS: @{v1} = len(set(@{e1}))
% % RuntimePrecond: type(@{e1}) == pandas.Series \end{minted}
% Another example of a trivial error is illustrated in Figure~\ref{}, where the precondition ensures semantic equality between the LHS and the RHS by explicitly evaluating both at runtime, thereby increasing computation overhead and rendering the rule impractical
 
\paragraph{Semantic Divergence due to Null Values} This category involves discrepancies between \pandas-specific implementations and native \python equivalents, particularly regarding the handling of missing data (\texttt{NaN} or \texttt{None}). An example of such a rule is shown below.

\begin{tcolorbox}[colback=backcolour, colframe=white, boxrule=0.1pt, arc=0.4mm, left=0.1mm, right=0.5mm, top=0.1mm, bottom=0.1mm]
    \textbf{LHS:} 
    \begin{minted}[fontsize=\small,breaklines,breakanywhere,bgcolor=lightgray!15, escapeinside=||]{python}
@{|\highlightast{Name}|: |\highlightvar{n1}|} = @{|\highlightast{Name}|: |\highlightvar{n2}|}.@{|\highlightast{Const(str)}|: |\highlightvar{c1}|}.min()

    \end{minted}
    \textbf{RHS:} 
    \begin{minted}[fontsize=\small,breaklines,breakanywhere,bgcolor=lightgray!15, escapeinside=||]{python}
@{|\highlightvar{n1}|} = np.min(@{|\highlightvar{n2}|}[@{|\highlightvar{c1}|}].values)
    \end{minted}

    \textbf{Runtime Preconditions:}
    \begin{minted}[fontsize=\small,breaklines,breakanywhere,bgcolor=lightgray!15, escapeinside=||]{python}
isinstance(@{|\highlightvar{n2}|}, pandas.DataFrame),
@{|\highlightvar{c1}|} in @{|\highlightvar{n2}|}.columns
    \end{minted}
\end{tcolorbox}

This rule is only correct when the input data does not have any NaNs. However, if the data does have NaNs, the output is different. While this error can be checked during the discovery stage to avoid the generation of such rules, we chose not to filter them. This is because these rules can be used in cases where the user is sure that the input data is clean and does not have any NaNs. However, since this is not true for \pandasbench, we excluded these rules from the final evaluation. 

\paragraph{Reference Errors} Some transformations achieve speedups by mutating DataFrames directly, bypassing the overhead of copying data. While they are safe in isolation, they can lead to incorrect behavior in rare corner cases due to data-flow dependencies across notebook cells.
An example of such a rule is shown below.
\begin{tcolorbox}[colback=backcolour, colframe=white, boxrule=0.1pt, arc=0.4mm, left=0.1mm, right=0.5mm, top=0.1mm, bottom=0.1mm]
    \textbf{LHS:} 
    \begin{minted}[fontsize=\small,breaklines,breakanywhere,bgcolor=lightgray!15, escapeinside=||]{python}
@{|\highlightast{Name}|: |\highlightvar{n1}|} = @{|\highlightast{Name}|: |\highlightvar{n1}|}.rename(columns={@{|\highlightast{Const(str)}|: |\highlightvar{c1}|}: @{|\highlightast{Const(str)}|: |\highlightvar{c2}|}})

    \end{minted}
    \textbf{RHS:} 
    \begin{minted}[fontsize=\small,breaklines,breakanywhere,bgcolor=lightgray!15, escapeinside=||]{python}
@{|\highlightvar{n1}|}.columns = [@{|\highlightvar{c2}|} if col == @{|\highlightvar{c1}|} else col for col in @{|\highlightvar{n1}|}.columns]
    \end{minted}

    \textbf{Runtime Preconditions:}
    \begin{minted}[fontsize=\small,breaklines,breakanywhere,bgcolor=lightgray!15, escapeinside=||]{python}
isinstance(@{|\highlightvar{n1}|}, pandas.DataFrame),
@{|\highlightvar{c1}|} in @{|\highlightvar{n1}|}.columns
    \end{minted}
\end{tcolorbox}
We chose not to exclude them from the final evaluation because these are corner-cases, and do not happen often in real-world programs. For instance, in \pandasbench, only 1 notebook out of 102 reported this error. In cases where the user intends to make an explicit soft copy of the dataset, they can chose to avoid these rules. 

% For instance, in Table~\ref{fig:speedupPerRule}, R1 circumvents copying overhead of the \texttt{rename} operation by setting the \texttt{inplace} parameter to True, achieving an average speedup of of $22.57\times$. However, if the target DataFrame is a \textit{view} (a slice or reference) of another DataFrame rather than an independent object, this transformations can propagate unintended changes to the source data.  

\end{document}